\documentclass{emulateapj}
\usepackage{apjfonts}
\usepackage{amsmath,amssymb,graphicx}
\usepackage{natbib}
\usepackage{graphics}
\usepackage{multirow}
\usepackage{amsbsy}
\usepackage{mathrsfs}
\usepackage[]{subfigure}

\def\bicep{{\sc Bicep}}

\def\bicepone{{\sc Bicep1}}
\def\biceptwo{{\sc Bicep2}}
\def\bicepthree{{\sc Bicep3}}

\def\spider{{\sc Spider}}

\def\keck{{\it Keck Array}}

\shorttitle{\biceptwo, \keck, and \spider\ Detectors} 
\shortauthors{\textsc{Ade et al.}}
\submitted{to be submitted to ApJ}

\begin{document}

\title{Antenna-coupled TES bolometers used in \biceptwo, \keck, and \spider}

\author{P.~A.~R.~Ade\altaffilmark{1}}
\author{R.~W.~Aikin\altaffilmark{2}}
\author{M.~Amiri\altaffilmark{3}}
\author{D.~Barkats\altaffilmark{4}}
\author{S.~J.~Benton\altaffilmark{5}}
\author{C.~A.~Bischoff\altaffilmark{6}}
\author{J.~J.~Bock\altaffilmark{2,7}}
\author{J.~A.~Bonetti\altaffilmark{7}}
\author{J.~A.~Brevik\altaffilmark{2}}
\author{I.~Buder\altaffilmark{6}}
\author{E.~Bullock\altaffilmark{8}}
\author{G.~Chattopadhyay\altaffilmark{7}}
\author{G.~Davis\altaffilmark{3}}
\author{P.~K.~Day\altaffilmark{7}}
\author{C.~D.~Dowell\altaffilmark{7}}
\author{L.~Duband\altaffilmark{9}}
\author{J.~P.~Filippini\altaffilmark{2,10}}
\author{S.~Fliescher\altaffilmark{11}}
\author{S.~R.~Golwala\altaffilmark{2}}
\author{M.~Halpern\altaffilmark{3}}
\author{M.~Hasselfield\altaffilmark{15}}
\author{S.~R.~Hildebrandt\altaffilmark{2,7}}
\author{G.~C.~Hilton\altaffilmark{12}}
\author{V. Hristov\altaffilmark{2}}
\author{H. Hui\altaffilmark{2}}
\author{K.~D.~Irwin\altaffilmark{13,14,12}}
\author{W.~C.~Jones\altaffilmark{15}}
\author{K.~S.~Karkare\altaffilmark{6}}
\author{J.~P.~Kaufman\altaffilmark{16}}
\author{B.~G.~Keating\altaffilmark{16}}
\author{S. Kefeli\altaffilmark{2}}
\author{S.~A.~Kernasovskiy\altaffilmark{13}}
\author{J.~M.~Kovac\altaffilmark{6}}
\author{C.~L.~Kuo\altaffilmark{12,13}}
\author{H.~G.~LeDuc\altaffilmark{7}}
\author{E.~M.~Leitch\altaffilmark{17}}
\author{N.~Llombart\altaffilmark{7}}
\author{M.~Lueker\altaffilmark{2}}
\author{P.~Mason\altaffilmark{2}}
\author{K.~Megerian\altaffilmark{7}}
\author{L.~Moncelsi\altaffilmark{2}}
\author{C.~B.~Netterfield\altaffilmark{5}}
\author{H.~T.~Nguyen\altaffilmark{7}}
\author{R.~O'Brient\altaffilmark{2,7,XX}}
\author{R.~W.~Ogburn~IV\altaffilmark{13,14}}
\author{A.~Orlando\altaffilmark{15}}
\author{C.~Pryke\altaffilmark{11}}
\author{A.~S.~Rahlin\altaffilmark{15}}
\author{C.~D.~Reintsema\altaffilmark{12}}
\author{S.~Richter\altaffilmark{6}}
\author{M.~C.~Runyan\altaffilmark{2,7}}
\author{R.~Schwarz\altaffilmark{11}}
\author{C.~D.~Sheehy\altaffilmark{17, 18}}
\author{Z.~K.~Staniszewski\altaffilmark{2}}
\author{R.~V.~Sudiwala\altaffilmark{1}}
\author{G.~P.~Teply\altaffilmark{2}}
\author{J.~E.~Tolan\altaffilmark{13}}
\author{A.~Trangsrud\altaffilmark{2,7}}
\author{R.~S.~Tucker\altaffilmark{2}}
\author{A.~D.~Turner\altaffilmark{7}}
\author{A.~G.~Vieregg\altaffilmark{17,18}}
\author{A.~Weber\altaffilmark{7}}
\author{D.~V.~Wiebe\altaffilmark{3}}
\author{P.~Wilson\altaffilmark{7}}
\author{C.~L.~Wong\altaffilmark{6}}
\author{K.~W.~Yoon\altaffilmark{13,14}}
\author{J. Zmuidzinas\altaffilmark{2,7}(for the \biceptwo, \keck, and \spider\ Collaborations)}

\altaffiltext{1}{School of Physics and Astronomy, Cardiff University, Cardiff, CF24 3AA, UK}
\altaffiltext{2}{Department of Physics, California Institute of Technology, Pasadena, CA 91125, USA}
\altaffiltext{3}{Department of Physics and Astronomy, University of British Columbia, Vancouver, BC, Canada}
\altaffiltext{4}{Joint ALMA Observatory, ESO, Santiago, Chile}
\altaffiltext{5}{Department of Physics, University of Toronto, Toronto, ON, Canada}
\altaffiltext{6}{Harvard-Smithsonian Center for Astrophysics, 60 Garden Street MS 42, Cambridge, MA 02138, USA}
\altaffiltext{7}{Jet Propulsion Laboratory, Pasadena, CA 91109, USA}
\altaffiltext{8}{Minnesota Institute for Astrophysics, University of Minnesota, Minneapolis, MN 55455, USA}
\altaffiltext{9}{Universit\'{e} Grenoble Alpes, CEA INAC-SBT, F-38000 Grenoble, France}
\altaffiltext{10}{Department of Physics, University of Illinois at Urbana-Champaign, Urbana, IL 61820, USA}
\altaffiltext{11}{Department of Physics, University of Minnesota, Minneapolis, MN 55455, USA}
\altaffiltext{12}{National Institute of Standards and Technology, Boulder, CO 80305, USA}
\altaffiltext{13}{Department of Physics, Stanford University, Stanford, CA 94305, USA}
\altaffiltext{14}{Kavli Institute for Particle Astrophysics and Cosmology, SLAC National Accelerator Laboratory, 2575 Sand Hill Rd, Menlo Park, CA 94025, USA}
\altaffiltext{15}{Department of Physics, Princeton University, Princeton, NJ 08544, USA}
\altaffiltext{16}{Department of Physics, University of California at San Diego, La Jolla, CA 92093, USA}
\altaffiltext{17}{Kavli Institute for Cosmological Physics, University of Chicago, Chicago, IL 60637, USA}
\altaffiltext{18}{Department of Physics, Enrico Fermi Institute, University of Chicago, Chicago, IL 60637, USA}

\altaffiltext{XX}{Corresponding author: rogero@caltech.edu}

\begin{abstract}
We have developed antenna-coupled transition-edge sensor (TES) bolometers for a wide range of cosmic
microwave background (CMB) polarimetry experiments, including \biceptwo,
\keck, and the balloon borne \spider.  These detectors have reached maturity and this paper
reports on their design principles, overall performance, and key challenges
associated with design and production.  Our detector arrays repeatedly produce
spectral bands with 20\%-30\% bandwidth at 95, 150, or 220~GHz.  The integrated antenna arrays synthesize symmetric co-aligned beams with controlled side-lobe levels.  Cross-polarized response on boresight is typically $\sim0.5\%$, consistent with cross-talk in our multiplexed readout system.  End-to-end optical efficiencies in our cameras are routinely 35\% or higher, with per detector sensitivities of $\mathrm{NET}$$\sim$300~$\mu\mathrm{K}_\mathrm{CMB}\sqrt\mathrm{s}$.  Thanks to the scalability of this design, we have deployed 2560 detectors as 1280 matched pairs in \keck\ with a combined instantaneous sensitivity of $\sim9~\mu\mathrm{K}_\mathrm{CMB}\sqrt\mathrm{s}$, as measured directly from CMB maps in the 2013 season.  Similar arrays have recently flown in the \spider\ instrument, and development of this technology is ongoing.
\end{abstract}


\section{Introduction}

Cosmic microwave background (CMB) polarimetry is a key observable to further our understanding
of cosmology in both the later and early universe.  Degree-scale B-mode
polarization can be used to constrain the tensor-scalar ratio $r$ and place
limits on the energy scale and potential form of inflation~\citep{Zaldarriaga_Seljak_CMBpol_1998, Kamionkowski_Kosowsky_CMBpol_1997}.
Arcminute B-mode measurements allow precise reconstruction of
the gravitational lensing potential at later times, which in turn can constrain the neutrino
masses and the dark energy equation of
state \citep{Kaplinghat_lensing}.  Precise measurements of E-mode polarization provide further cosmological information on the plasma physics at recombination.  
Very precise future lensing polarization maps could ultimately be used for deeper searches of inflationary polarization.   Finally a future space mission has the potential to measure large-scale polarization to astrophysical limits, for precise tests of inflation.  However, these ambitious scientific goals require high sensitivity cameras with exquisite control of systematic errors. 

To meet this need, we have developed large arrays of dual-polarized
antenna-coupled transition edge sensor (TES) bolometers.  The key
feature in our design is optical coupling through a planar antenna, allowing the entire design to be fabricated with scalable
photolithographic techniques.  This has allowed us to rapidly deploy arrays for \biceptwo, \keck, and the balloon borne \spider,
amounting to over 6000 detectors fielded as of this writing.  
We deployed \biceptwo\ for observing in 2010, three \keck\ 150~GHz cameras for observing in 2011, and five 150~GHz cameras for \keck\ in 2012.  In 2014, we replaced two \keck\ 150~GHz cameras with 95~GHz cameras, and in 2015 we replaced two more with 220~GHz cameras (fielding an additional 1000 detectors).  
\spider\ recently conducted a long-duration flight from McMurdo Station in early, deploying an additional 2400 detectors~\citep{2013JCAP...04..047F}.

This paper describes the design principles of these detectors, as well as their on-sky performance and
some challenges associated with their production.  We draw from our extensive experience
with the aforementioned experiments, and describe the as-deployed performance
of our devices.

The format of this paper is as follows.  In \S\ref{sec:camera} and \S\ref{sec:design}
 we describe the cameras and detector element designs, elaborating beyond the material in the \biceptwo~Instrument Paper \citep{b2instrument}.  \S\ref{sec:fab} describes the fabrication techniques used.  To optimize our designs and recipes, we had to characterize the optical properties of our thin films for millimeter waves, which we describe in \S\ref{sec:film_properties}.  \S\ref{sec:beams} describes the antenna array beam synthesis as well as related processing and design challenges.  Our high on-sky detector yield is only possible through control of detector properties across each tile, which we discuss in \S\ref{sec:array_prop}.  \S\ref{sec:sensitivity} describes the attained camera sensitivity and we show rough agreement between those measurements and a noise model.  Finally, \S\ref{sec:conclusions} offers some concluding remarks and describes future endeavors.

\section{Overview}
\label{sec:camera}

\begin{figure*}[ht]
\centering
\subfigure{
\includegraphics[width=2\columnwidth]{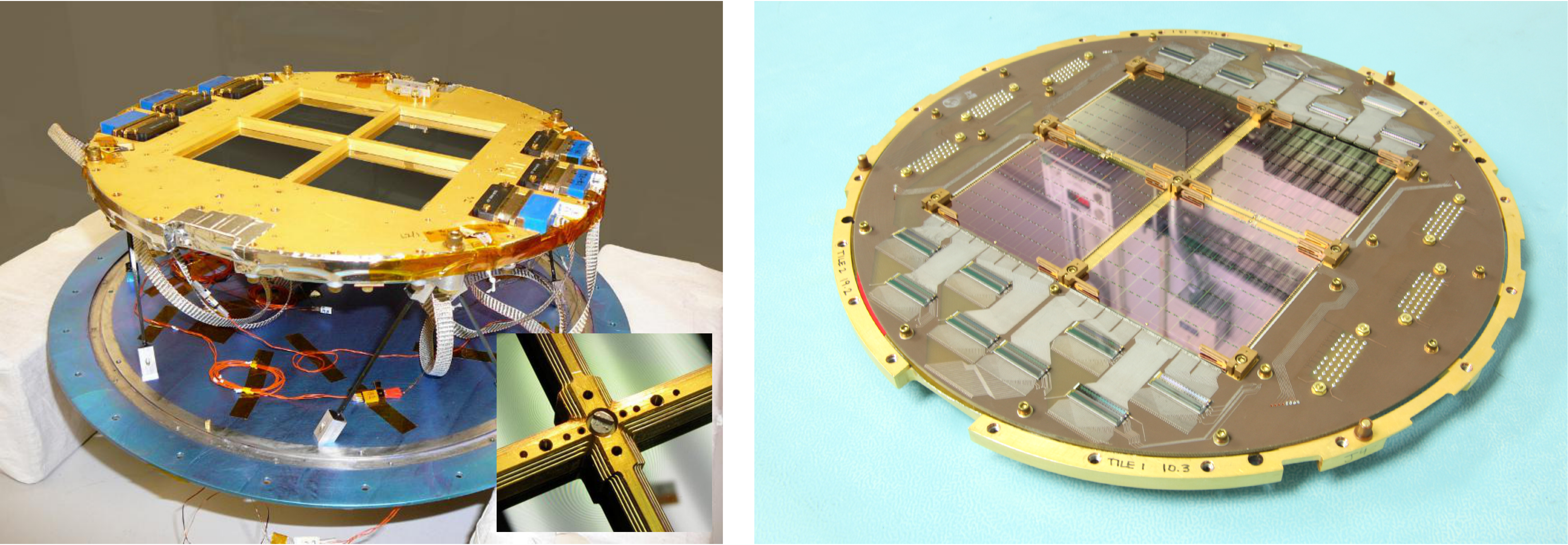}
}
\subfigure{
  \includegraphics[width=2\columnwidth]{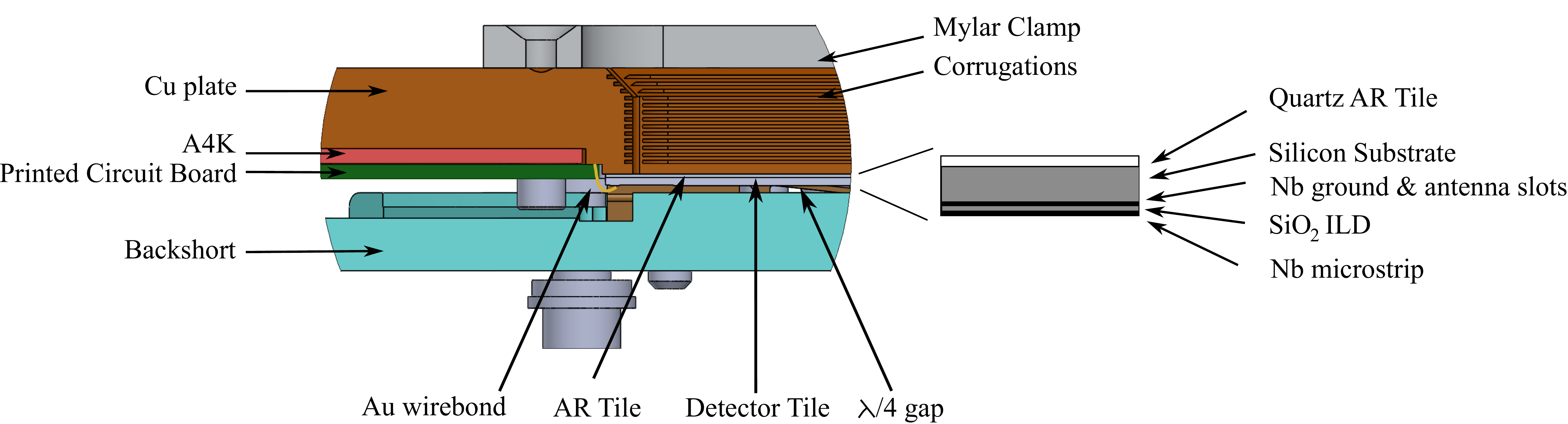}
}
\caption{Focal Plane pictures and cross-section.  {\it Upper}: Top (left) of the \biceptwo\ focal plane and bottom with backshort removed (right).  The arrays of 64 detector pairs per tile are visible at top right.  {\it Lower}: Major component layers of the focal plane design, with an expanded view of the tile layers at right.  Gold wire-bonds thermally sink the tiles to the frame and the frame bears corrugations to suppress coupling to the detectors.  Only the detectors at the tile perimeter are at risk of frame coupling.  Corrugations are visible in the inset photo in the upper left panel.\label{fig:FPU_pics_stratification}}
\end{figure*}

\paragraph{Cameras} Because the detector arrays are coupled to experiments with refracting optics, we briefly summarize their architecture.  The \biceptwo, \keck, and \spider\ cameras are similar $f$/2.2 26~cm aperture telecentric refracting telescopes, which re-image the sky onto the focal plane with pairs of high-density
polyethylene (HDPE) lenses.  \biceptwo\ and \keck\ contain teflon, nylon, and reflective metal-mesh edge-filters that filter incident radiation to maintain low thermal loading on the focal plane such that
internal closed-cycle $^\mathrm{3}$He/$^{\mathrm3}$He/$^\mathrm{4}$He sorption fridges can cool the focal plane to $280$~mK \citep{duband99}.  Due to loading constraints at float, the \spider\ cameras cannot employ lossy teflon filters and instead utilize metal-mesh filters.  They still use a nylon filter on the cold side of 4K.  \spider\ uses a $\sim 2$~K pumped helium bath and single stage closed cycle $^\mathrm{3}$He fridges to chill its focal planes to $\sim300$~mK.  \spider\ uses ultra-high-molecular-weight polyethylene (UHMWPE) windows in its low pressure flight environment, whereas \biceptwo\ and \keck\ use  closed cell nitrogen filled Zotefoam (respectively Propozote PPA30 and Plastazote HD30).  We refer the reader to the Instrument Paper for more \biceptwo-specific details \citep{{b2instrument}}.

\paragraph{Focal plane} Each focal plane (Figure~\ref{fig:FPU_pics_stratification}) contains four detector tiles mounted to a common copper
frame.  Each tile is a square of silicon cut from a standard 100~mm wafer, typically 550~$\mu$m thick.  The 150~GHz \biceptwo, \spider, and \keck\ focal planes contain four tiles, each consisting of 64 dual orthogonally polarized detector pairs, providing 512 detectors per camera.  The detectors' beams terminate on their camera's cold stop at -15~dB of the main lobe with $1.8f\lambda$ spacing.  The 95~GHz \keck\ cameras have tiles containing 36 orthogonally polarized detector pairs, and thus 288 detectors per camera.  These detectors' beams terminate on their camera's stop at -12~dB of the main lobe with $1.5f\lambda$ spacing.
Current 220~GHz focal planes contain four tiles of 64 detector-pairs each in order to match a readout system 
designed for the 150~GHz cameras; this under samples the focal plane ($2.6f\lambda$ spacing), which will be addressed with 
a future higher density focal plane design.  The 220~GHz detectors' beams are similar in size to the 150~GHz beams, terminating on the stop at -15~dB.

We mount the tiles to the aforementioned gold-plated copper frame and 
thermally sink the tiles
with gold wire bonds linking the frame to gold bond-pads that directly contact
the silicon substrate of the tile.  The antennas are patterned on the non-illuminated side of the wafer, so that light arrives 
at the antenna through its silicon substrate (see \S \ref{sec:antennadesign} for elaboration).  A quartz anti-reflection tile is mounted on the illuminated side of the wafer, 
while the antenna side faces a superconducting niobium reflective back-short placed $\lambda/4$ away.  Testing of 
early \biceptwo\ prototype focal planes suggested that near the tile edge, antennas for one of the two polarizations 
may couple to the frame, yielding elliptical beams in the far field.  We have mitigated this coupling in all deployed focal planes by keeping the edge detector pairs a few wavelengths away from the frame and by corrugating the frames with specifically
chosen depth and impedance grooves, optimized with simulations performed in the CST Microwave Suite \citep{CTS_manual}.

\paragraph{Polarized Detector Elements} Our detector design is entirely planar and does not require horns or other contacting optics.   In each detector element, optical power couples to two co-located, orthogonally polarized  planar antenna arrays, each composed of slot sub-radiators patterned in a superconducting niobium (Nb) ground plane.  All slots of a given orientation are coherently combined through a microstrip summing tree to synthesize a single equivalent antenna for that polarization orientation.  Power from each antenna is passed through an on-chip band-defining filter before being dissipated on a suspended bolometer island.  A superconducting TES on that island detects variations in the power received by the antennas.

\section{Detector Design}
\label{sec:design} 

\subsection{Antenna Design}
\label{sec:antennadesign}

\begin{figure}[ht]
\centering
\subfigure{
\includegraphics[width=1\columnwidth]{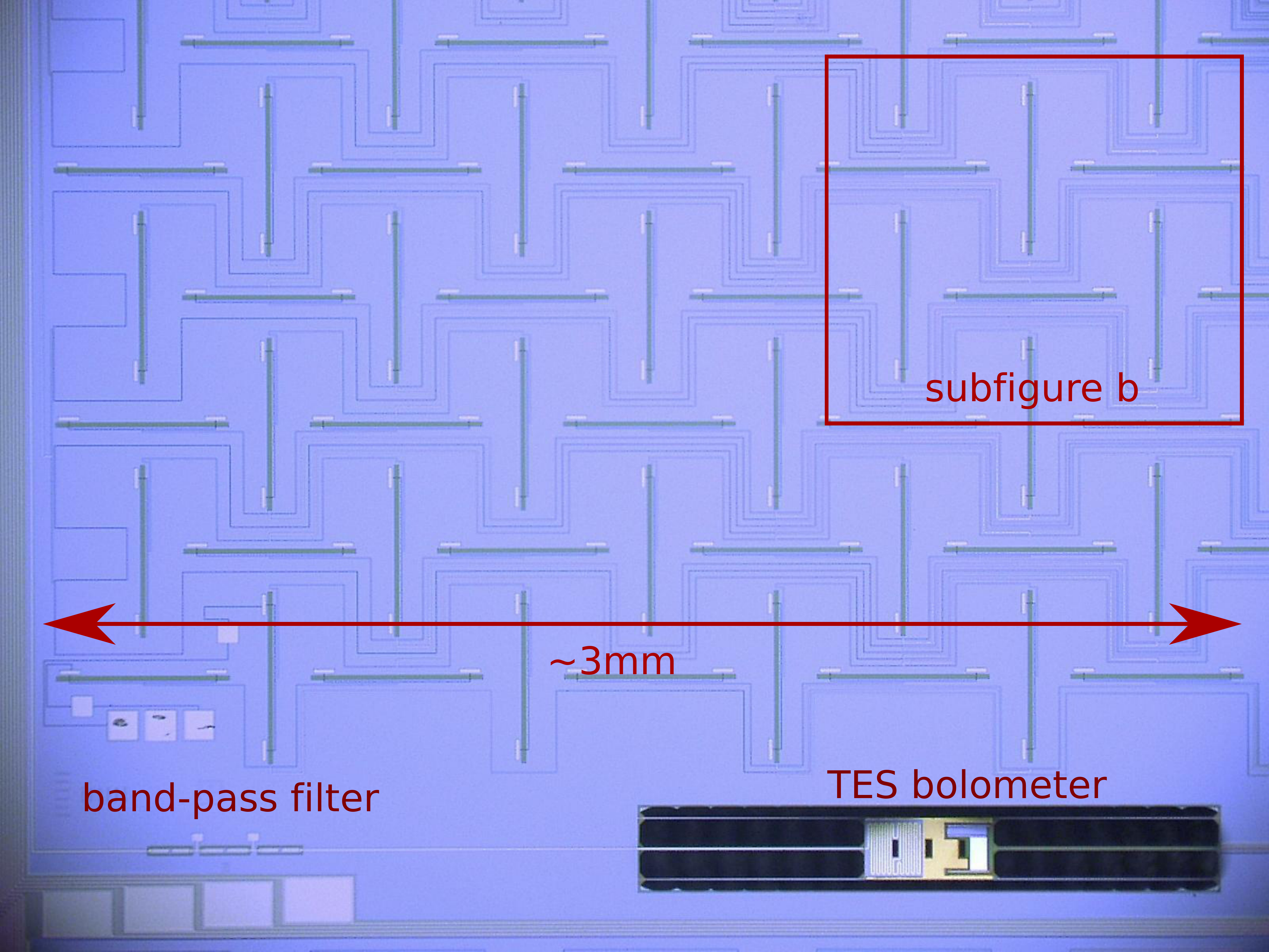}
}
\subfigure{
  \includegraphics[width=1\columnwidth]{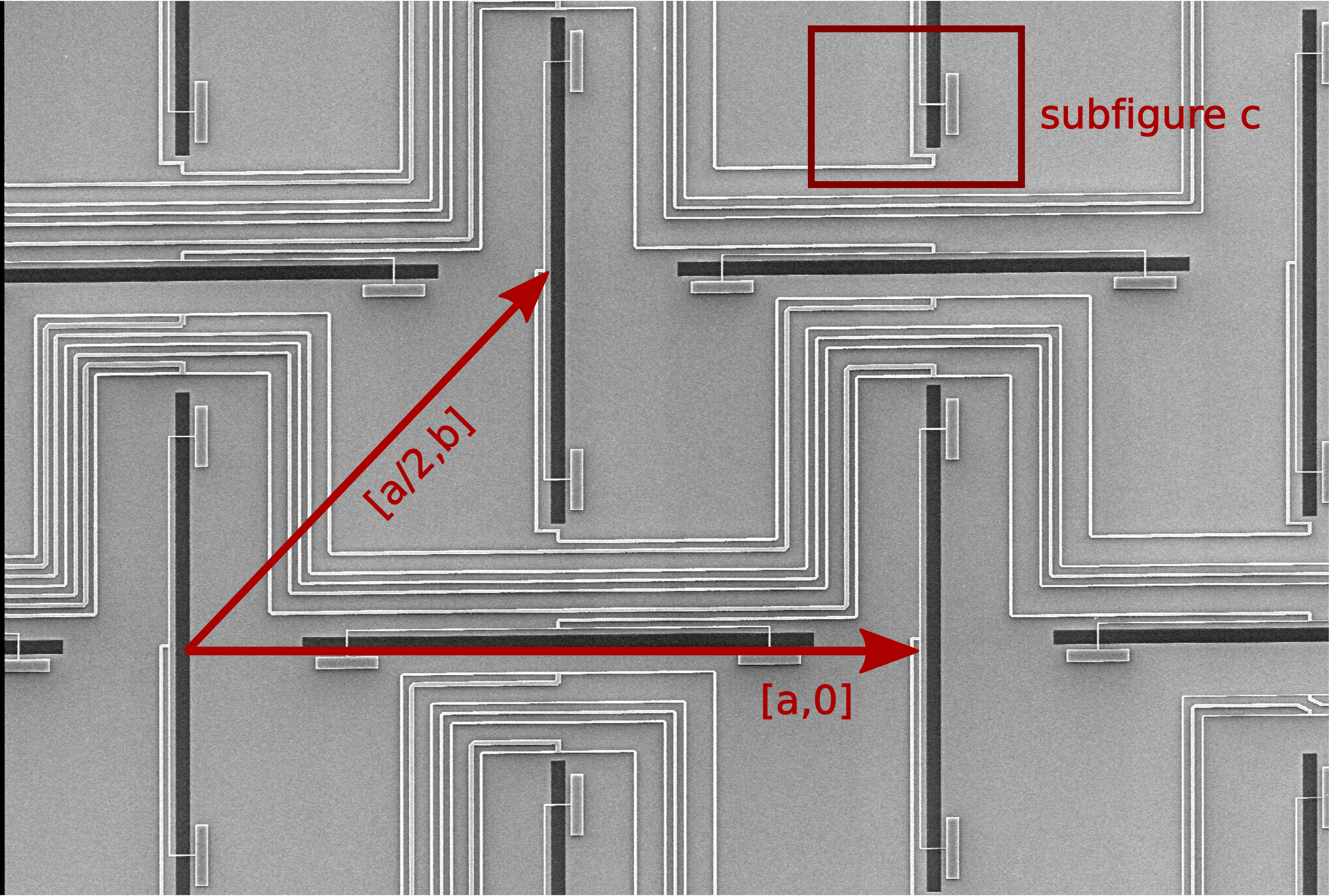}
}
\subfigure{
  \includegraphics[width=1\columnwidth]{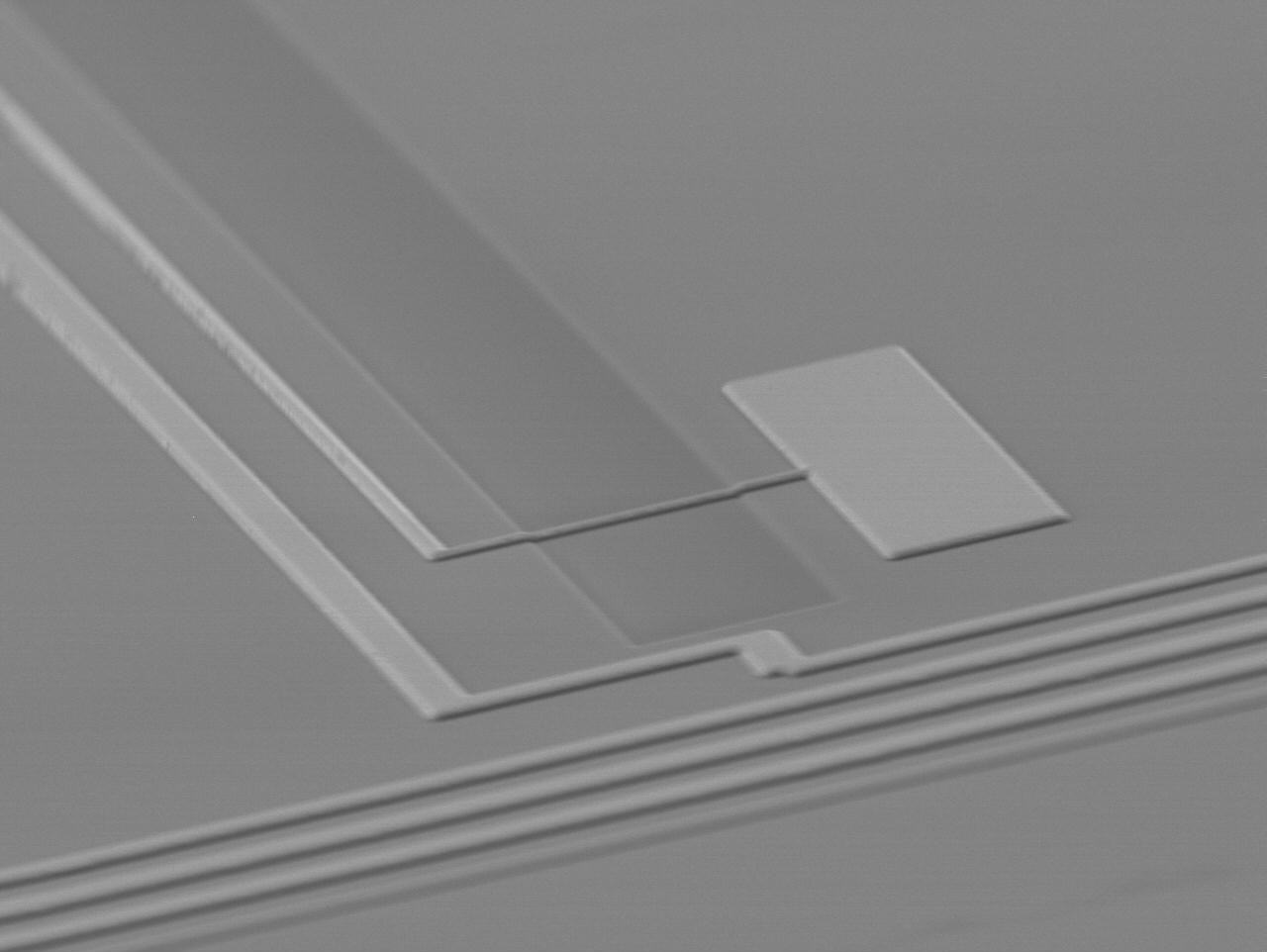}
}
\caption{Optical and SEM photographs of major features of the antenna array. {\it Upper}: one quarter of a detector element.  The antenna array, one filter, and one TES bolometer are visible, as well as DC readout lines for the detector.  {\it Middle}: SEM micrograph of the slot array (dark rectangles) and oblique Bravais lattice (arrows).  The thin white lines comprise the microstrip feed.  For 150~GHz detectors, a$\sim600~\mu\mathrm{m}$ and $b=a/2\sim300~\mu\mathrm{m}$; the slot dimensions and spacing in the 95~GHz are 63\% larger and those in the 220~GHz detector elements are 47\% smaller.  {\it Lower}: SEM micrograph of microstrip crossover and shunt capacitor at a sub-antenna slot.\label{Pixel_Slot_array_pics}}
\end{figure}

The antenna slots in each detector must be spaced to Nyquist sample the focal plane surface to avoid grating lobes that would rapidly change the impedance with
frequency \citep{2008SPIE.7020E..1IK}.  The antenna pattern of each axis of an array is calculated from the $N$ elements per linear
dimension spaced at distance $s$ as follows:
\begin{eqnarray}
 A(\theta) &=&\sum_{m=-(N-1)/2}^{(N-1)/2} e^{-j 2 \pi \frac{m~s~\sqrt{\epsilon_r}}{\lambda_o}\sin(\theta)}  \nonumber \\
 &=&\frac{\sin(N \pi s\sqrt{\epsilon_r}\sin\theta/\lambda_o)}{\sin( \pi s\sqrt{\epsilon_r}\sin\theta/\lambda_o)} ,
 \label{sinc_dist}
\end{eqnarray}
where $\lambda_o$ is the free-space wavelength, $\epsilon_r$ the relative permittivity of the surrounding medium, and the sum is across sub-antennas indexed by $m$ .  In addition to the strong peak in the normal direction ($\theta=0$), there are
grating lobe peaks when $\sqrt{\epsilon_r}s \sin(\theta)/\lambda_o$ is a
positive integer.  To avoid these lobes, the slot spacing must be
\begin{equation}
s \le \frac{\lambda_{o,\mathrm{min}}}{\sqrt{\epsilon_r}}\left( 1-\frac{1}{N}\right),
\label{grating_lobe_condition}
\end{equation}
where $\lambda_{o,\mathrm{min}}$ is the minimum wavelength of operation and the term
in parentheses accounts for the finite width of the grating-lobe peaks.  For
the 150~GHz detectors fabricated on silicon ($\epsilon_r=11.8$) with an upper
band edge of 180~GHz ($\lambda_{o,min}=1.7~\mathrm{mm}$), the spacing must satisfy $s\le460~\mu\mathrm{m}$.
To achieve this high density of radiators without intersection, we couple
power to offset (echelon) pairs of radiators with a Bravais lattice defined
in Figure~\ref{Pixel_Slot_array_pics}.  These dual-slot sub-arrays tile the detector elements
in $8\times 8$, $10\times 10$, or $12\times 12$ versions, where the overall detector size is chosen
to match the $f$/2.2 camera optics.  By using such a large number of sub-antennas, we avoid excessive excitation of substrate modes that
might degrade the detector's efficiency.  The offset slot-pair geometry allows the two orthogonally polarized 
antenna arrays within each detector pair to be co-located (visible in upper two panels of Figure \ref{Pixel_Slot_array_pics}).

It is energetically favorable for the antennas to receive power through
the silicon substrate to the vacuum side at a ratio of $377~\Omega:\lvert Z_{\mathrm{Si}}^{\mathrm{in}}\rvert$, where $Z_{\mathrm{Si}}^{\mathrm{in}}$ (in equation~\ref{green_function})
differs from silicon's TEM impedance of $110~\Omega$ in phase because
standing waves in the substrate modify the effective impedance seen by the slots.
We exploit this power difference by orienting the tiles with the silicon substrate side towards the sky and then terminating the back response on the vacuum side with a
$\lambda$/4 back-short.  We mount $\lambda$/4 quartz anti-reflection (AR) tiles
to minimize reflection at the air-substrate interface.

\subsection{Antenna Impedance}

Optical power couples from the slots to a microstrip feed network. To minimize return loss at this interface, we need to accurately compute the input impedance of that feed network. We have written custom software that provides increased speed and versatility for this task over commercial options.

We compute the input impedance of our antenna from the required jump-discontinuity
in magnetic fields across the slot at the microstrip feed-points
($J_{\mathrm{feed}}=\Delta H_x$) where we impose a driving current $J_y$.
Modeling the currents in the ground plane with fictitious magnetic currents
${\bf M}=\hat{n}\times {\bf E}$ running longitudinally down the slot, 
the field jump-discontinuity requires:
\begin{equation}
 \begin{array}{rclcl}
  J_y&=& &\int& dx' dy' {\bf M}(x',y')\\
   &&&& \left[G_{H_{1x},M_x}(x-x',y-y') - G_{H_{2x},M_x(x-x',y-y')}\right]
   \label{jump_continuity_eqn}
 \end{array}
\end{equation}
The antenna impedance can be inferred once these currents ${\bf M}$ are solved for.
The Green functions $G_{H_{nx},M_x}$ describe the fields on either side of the
slot ($n=\left[1,2]\right]$) resulting from an infinitesimal magnetic current in the {\bf x} direction (parallel to the slot.)
In the spectral domain, these have the form
\begin{equation}
\tilde{G}_{H_xM_x}(k_x,k_y,z=0)=\frac{\cos(\phi_k)}{Z^\mathrm{in}_\mathrm{TM}(k_{\rho})}+\frac{\sin(\phi_k)}{Z^\mathrm{in}_\mathrm{TE}(k_{\rho})}
\label{green_function}
\end{equation}
where $\phi_k=\arctan{(k_y/k_x)}$ and $k_{\rho}=\sqrt{k_x^2+k_y^2}$  \citep{1133646}.  The dielectric films support transverse electric (TE) and transverse magnetic (TM) modes with different impedances.  The impedance $Z^{\mathrm{in}}_{\mathrm{TE,TM}}$ of each mode ``seen'' by the slots can be computed by solving for the standing waves of electric and magnetic fields in a manner analogous to treating the silicon substrate and quartz AR coating as transmission lines and then transforming the impedance of free space ($377~\Omega$) through them. \citep{1133646}  The backshort (ground) transforms through a quarter-wave space, presenting a parallel impedance that is open at the band center.   The silicon substrates are a standard 100~mm diameter $550~\mu\mathrm{m}$ thick wafer, and we optimized the 95~GHz and 150~GHz impedance matching with this thickness in mind.  The 220~GHz versions are fabricated on thinner $370~\mu\mathrm{m}$ thick silicon substrates to scale this design.

We expand the unknown fictitious currents $M(x,y)=\sum_iV_i F_i(x,y)$ in a one-dimensional piecewise sinusoidal basis \citep{1143638}:
\begin{equation}
F_i(x,y)=\frac{1}{\pi\sqrt{(w/2)^2-y^2}} \begin{cases}
    \frac{\sin(k_{\rho}(\Delta x-|x-x_i|))}{\sin(k_{\rho}\Delta x)},& |x-x_i|\leq \Delta x\\
    0,              & \text{otherwise}
\end{cases}
\end{equation}
localized in physical space at position $x_i$ over the slot, with sections of length $\Delta x$ and width $w$ corresponding to the slot width.  This basis models the expected current distribution on the shorter slot dimensions,
leaving the unknown distribution only on the longer dimension. Moreover the Fourier transform of the basis function has a compact analytic
form.  These features greatly simplify the computations.  Solving
Equation~\ref{jump_continuity_eqn} with a Galerkin moment method reduces it to a
matrix equation $[Y_{ij}][V_i]=[I_j]$ \citep{1143638}, where
\begin{equation}
Y_{ij}=\int dk_x dk_y \tilde{F}^*_i(k_x,k_y)\tilde{G}(k_x,k_y,z=0)\tilde{F}_j(k_x,k_y)
\label{admittance_eqn}
\end{equation}
The inverse of the admittance matrix in Equation~\ref{admittance_eqn} is an
impedance between basis functions.  The self and mutual impedance between
basis functions located at feed-points is the input impedance that our
feed-lines must match.

\begin{figure}[htbp]
\centerline{\includegraphics[width=.8\columnwidth]{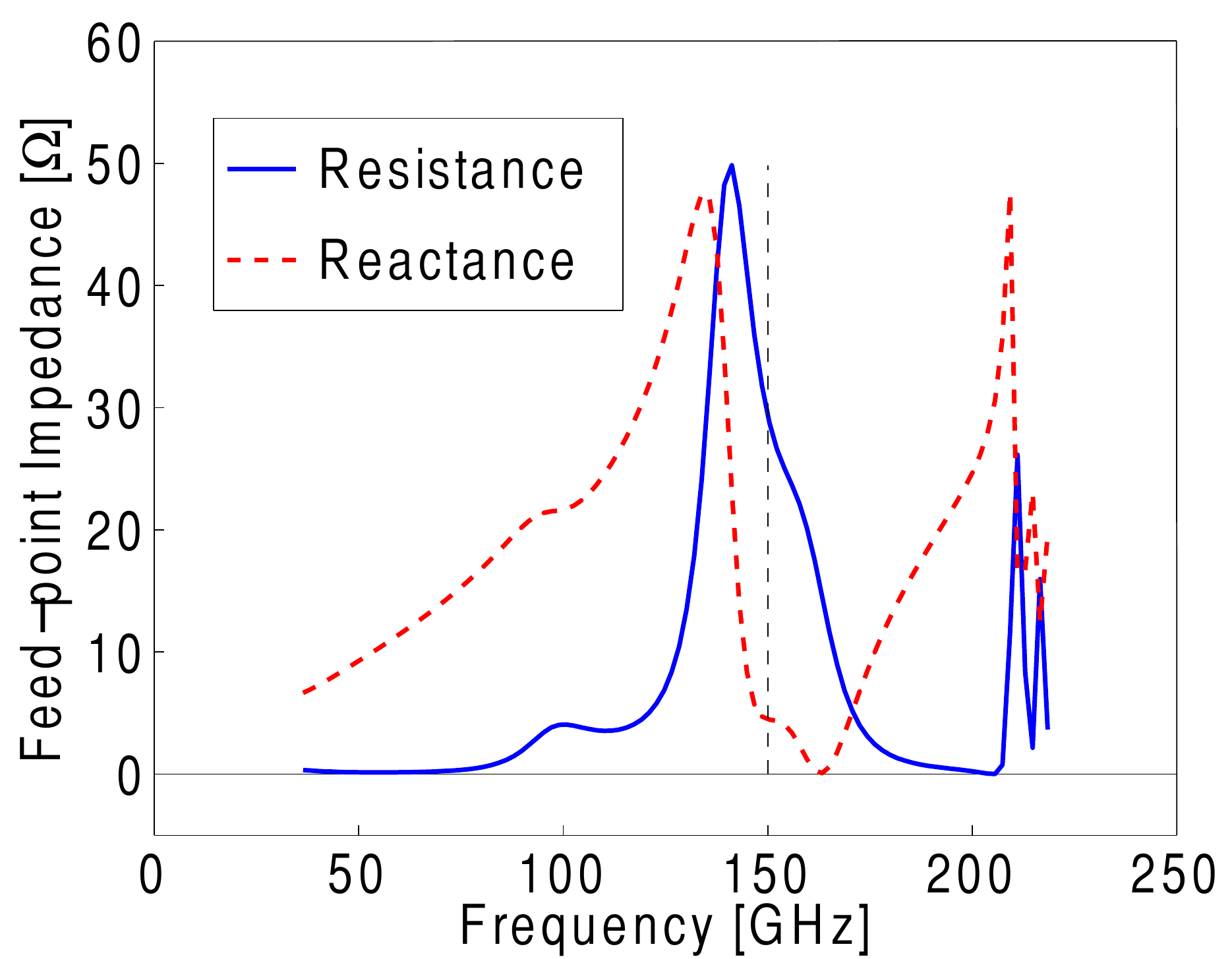}}
\caption{Simulated feed-point antenna impedance vs. frequency. The dashed vertical line is the band center.\label{Impedance_v_f}}
\end{figure}

We avoid complicated numerical integration and further reduce the number of
unknown $V_i$s by assuming the arrays are infinite in extent and applying
periodic boundary conditions where the antenna properties repeat for spectral
translations along the lattice vectors:
\begin{eqnarray}
k_x^{n}&=&\frac{2 \pi n}{a} \nonumber \\
k_y^{m,n}&=&2 \pi \left( \frac{n}{a}+\frac{m}{b}\right)
\end{eqnarray}
where $a$ and $b$ are the distances between the slots in the horizontal and vertical axes (see Figure~\ref{Pixel_Slot_array_pics}).  This approximation is valid for slots in the interior of the array and reduces the integrals to a discrete sum that we terminate at a wavevector
high enough to allow convergence.  Figure~\ref{Impedance_v_f} shows the computed
input impedance against frequency for the 150~GHz antennas.  The oscillatory behavior at 200~GHz is expected and real ({\it i.e.} not a numerical artifact) and coincides with the onset of grating lobes.  We note that we have performed more numerically intensive simulations of finite slot arrays where continuous integration retains edge effects to some degree and these agree well with the infinite array approximation.

We couple power to microstrip lines at the points near the slot ends where the lines cross and shunt to the ground-plane on the opposite side (see lower panel of Figure \ref{Pixel_Slot_array_pics}).  This coupling effectively transmutes the electric fields across the slot into fields between the microstrip conductor and ground plane, provided that impedances are well matched.  To avoid grating lobes, we keep the slots as short as possible and excite the slots at their first resonance.  However, the radiation resistance of center-fed slots at this resonance is nearly \mbox{300~$\Omega$}, caused by a current node at the slot center.  By symmetrically feeding the slots with a pair of off-center microstrip lines close to the current anti-nodes, we greatly reduce the radiation resistance to $\sim40~\Omega$.  The microstrip feed lines must match this real antenna resistance, which, for our dielectric and metal film thicknesses, corresponds to the thinnest lines that we can reproducibly fabricate.  The feed points also have a $5~\Omega$ inductive reactance, which is relatively stable over the band.  We tune this away with series capacitors that shunt current to ground (also visible in Figure \ref{Pixel_Slot_array_pics}).  Future designs may use more complicated terminations ({\it e.g.}, open stubs) in lieu of capacitors to expand the bandwidth, allowing for co-located dual-color detectors.

\subsection{Microstrip feed}

\begin{figure}[htbp]
\centerline{\includegraphics[width=.8\columnwidth]{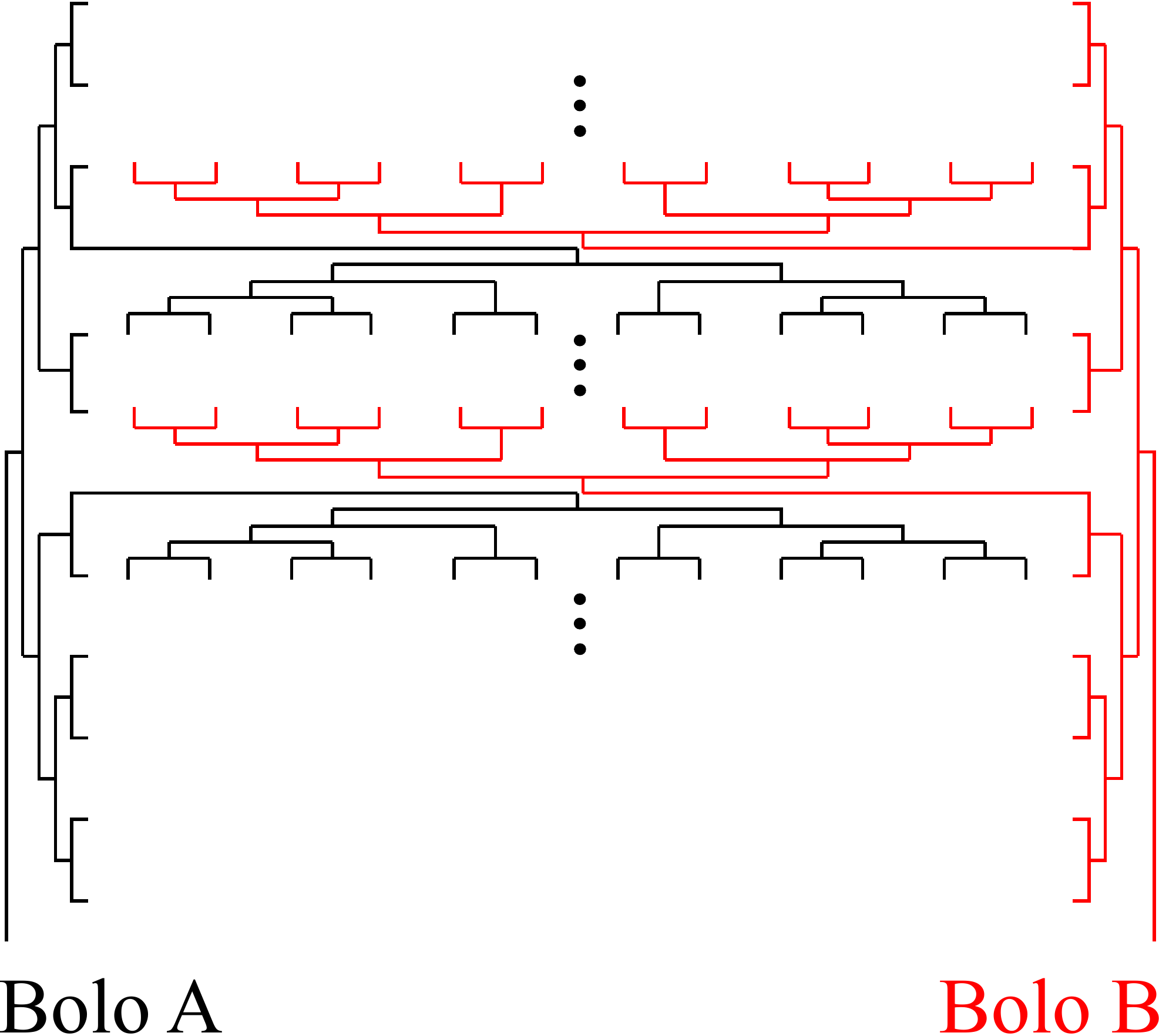}}
\caption{Abbreviated feed-network schematic for one detector pair, showing the branch structure for 
the two polarization summing trees. \label{Summing_tree_topology}}
\end{figure}

The waves from the slots coherently sum in the microstrip
feed-network, which accomplishes beam synthesis in lieu of a horn.  Figure~\ref{Summing_tree_topology} 
summarizes this topology.
This network combines waves across rows, and then sums the waves from
each row in a column tree at the side of the detector element, presenting a single
microstrip line in each polarization.  The two orientations of slots
couple power to two independent feeds that tightly interlock across
the detector pair.  The feeds are meant to be corporate, combining all waves
with uniform phase, but the trees are not necessarily binary.
In fact, the 220~GHz and 150~GHz detectors have a $12\times 12$ array
format while the 95~GHz ones are $10\times 10$, neither of which
is a power of 2.

Waves sum in microstrip tee-junctions, with impedances chosen to match across each junction when looking from the port closest to the bolometer.  We pick the ratio of impedances on the ports closest to the slots to determine the illumination pattern. To accomplish this synthesis
with low return loss, we must construct microstrips with correct impedances,
which requires accurate knowledge of dielectric constants and the
 penetration depth into our superconducting niobium films.  (See \S\ref{sec:film_properties}.)  Thus far, all detectors deployed for \biceptwo\, \keck\, and \spider\ produce a uniform top-hat illumination where the power splits
are in proportion to the number of slots on each side of the junctions.
The microstrip impedances can be chosen to synthesize an arbitrary
pattern, however, and Gaussian tapered feeds have recently been deployed in the \bicepthree\ camera.

Coupling between microstrip lines can generate phase
errors across the antennas, which we elaborate on in
\S\ref{sec:beams}.  The present designs have judiciously
spaced the adjacent lines to minimize this coupling.  To negate any
residual phase errors, the feed-networks also contain sections of
transmission line before each slot whose length we can vary
during fabrication.

\subsection{Band-defining filters}
\label{sec:filters}
\begin{figure}[htbp]
\centerline{\includegraphics[width=.8\columnwidth]{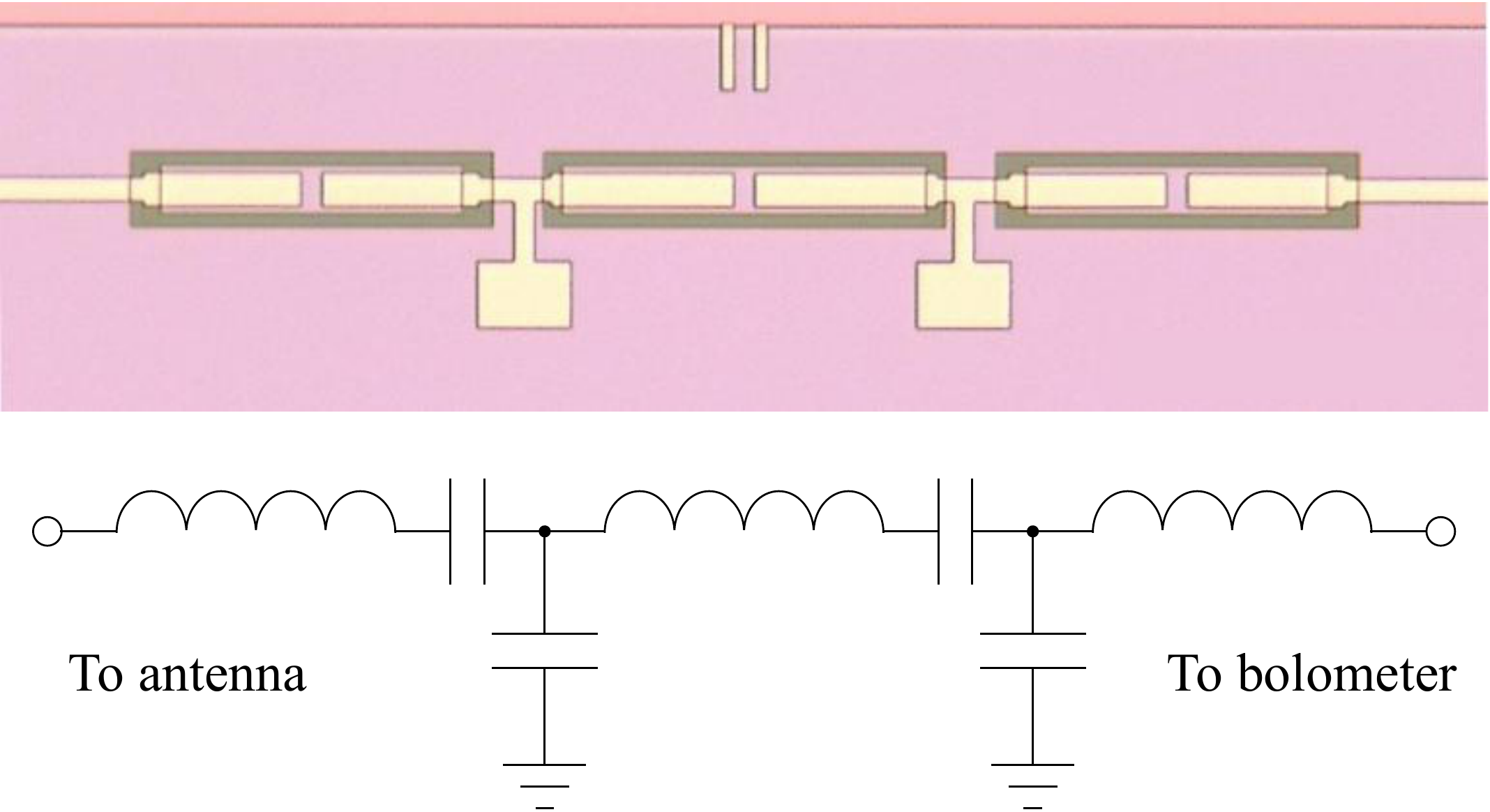}}
\caption{Microscope photograph of filter and equivalent circuit for 150~GHz. \label{fig:filter_circuit}}
\end{figure}

Each microstrip feed contains an integrated band-defining filter
between the antenna feed and bolometer.  We use a three-pole design as a compromise between bandwidth 
and loss: additional poles would let us increase the bandwidth and still avoid atmospheric features, but the 
additional passes of the waves through each resonator increase filter loss.  
Figure~\ref{fig:filter_circuit} shows a micrograph of the filter and a schematic of its equivalent circuit.  
We realize the resonators with lumped components, which do not suffer from the high frequency resonant 
leaks present in $\lambda/2$ and $\lambda/4$ transmission line resonators.

Each pole is a series LC resonator, in which the inductors are short stretches
of high-impedance coplanar waveguide (CPW).  The CPW impedance $Z=\sqrt{L/C}$
of roughly $50~\Omega$ exceeds the surrounding lines and acts as a series
inductor by allowing strong magnetic fields in the CPW ground-plane gaps.
This inductance is almost entirely magnetic; kinetic inductance from the
niobium superconducting microstrip and ground layers only makes a minor correction.  The series capacitors are
parallel-plate metal-insulator-metal between upper and lower niobium films
using the microstrip SiO$_2$ as the dielectric.  In an ideal design the central
resonator would be a parallel LC resonator shunting to ground.  For
ease of fabrication we instead build a series resonator and invert its impedance 
on resonance using shunt capacitors to ground and reduced series capacitors~\citep{4545888}.

We numerically optimize the filter in simulations with the commercially
available Sonnet software \citep{Sonnet_manual}.  The optimizations are constrained to maintain
an entrance impedance of $10~\Omega$ which results in fabricable capacitors and inductors.
Our design results in inductors longer
than $\lambda/8$ that have some slight parasitic shunt capacitance, but this
does not degrade the performance appreciably.  The filter in
Figure~\ref{fig:filter_circuit} is tuned for a 150~GHz band-center.  Filters for
95~GHz and 220~GHz versions have components with the same reactance at the
resonant frequencies.  We show spectra and summarize spectral response features in \S\ref{sec:array_prop}.

\subsection{Bolometer Design}
\label{sec:det_design}

We terminate and thermalize millimeter-wave power on a released bolometer
island in a meandered lossy gold microstrip.  150~GHz waves are constrained
to propagate through a skin depth of $180$~$\mathrm{nm}$, and over a length of several wavelengths
the line absorbs 99\% of the power between two passes.
We scale the 220~GHz and 95~GHz designs appropriately to maintain $-20$~dB return loss.  

\begin{figure}[htbp]
\centerline{\includegraphics[width=1\columnwidth]{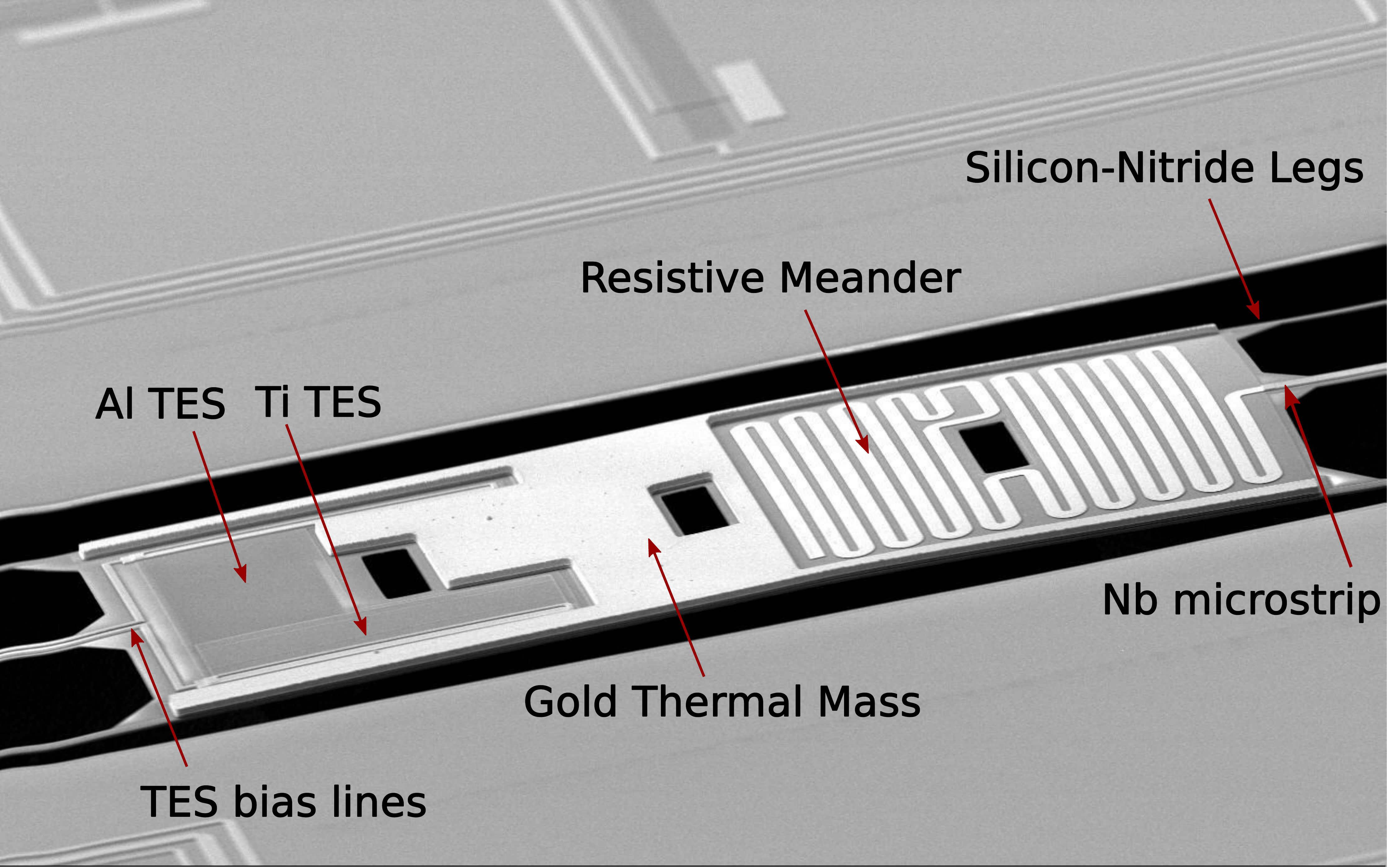}}
\caption{Electron micrograph of a released TES bolometer, illustrating its major components.  
The gold-meandered microstrip termination is at the right of the photograph and the TESs at left.  
The thicker gold film in the center of the island ensures thermal stability.  \label{TES_pic}}
\end{figure}

This termination is in close thermal contact with two transition edge sensors:
an aluminum TES with a transition temperature $T_c\sim1.2~\mathrm{K}$ for lab tests and a $\sim 60~\mathrm{m}\Omega$ titanium TES with a $T_c\sim0.5~\mathrm{K}$ for on-sky observations. The saturation power (optical plus electrical power) needed to bring the TES temperature to $T_c$ is

\begin{equation}
P_{sat} =G_c T_c \frac{1-(T_o/T_c)^{n+1}}{n+1}
\label{Psat}
\end{equation}
where the conductance $G_c$ is evaluated at the transition temperature $T_c$ and the surrounding heat bath is $T_o\sim280~\mathrm{mK}$.  The exponent
$n$ reflects the thermal carriers in the legs, where $n=1$ would
correspond to electrons and $n=3$ would correspond to phonons in 3-D materials.  Our typical devices are described by $n=2.5$.  The aluminum $T_c$ affords the bolometer
a high saturation power for use under a 300~K background, but has higher noise that would be unacceptable for astrophysical observations.  

Each bolometer island is suspended from the tile by six isolation legs: one carrying the 
microstrip from the antenna, one carrying the TES DC lines, and four thinner legs for 
mechanical stability.  Some newer devices use four legs, while \spider\ employs long meandered
legs as described below.  We tune the bolometer legs to
achieve the design thermal conductance.  We aim
to keep the phonon noise subdominant to the photon noise,
but also to keep the $G$ high enough that the detectors do not saturate under typical
on-sky loading conditions, which for \biceptwo\ is $4$--$6~\mathrm{pW}$.  We typically build in a safety factor of 2 beyond the expected
maximum loading; for \biceptwo\ this resulted in $G_c=80$--$150~\mathrm{pW/K}$, as described in \S\ref{sec:array_prop}.   Two detector pairs in the corners have the microstrip opened between the antenna and bolometers such that the TESs do not receive optical stimulation through the microstrip feed.  As described in \S\ref{sec:direct_stim}, we use these four bolometers for diagnostic measurements and to monitor direct stimulation of our detectors by optical power that bypasses the microstrip circuitry and may thus not have the required spectral and polarization properties.

Thick evaporated gold (visible in Figure~\ref{TES_pic}) is added to the bolometer islands to
 boost the heat-capacity by $C_{\mathrm{Au}}\simeq0.5-0.3~\mathrm{pJ/K}$, bringing 
the time constant $\tau=C/G$ to around a millisecond.  We infer this from
responses to a 1~Hz square-wave modulated broad-spectrum noise signal.  The
additional gold is important because of the strong electrothermal feedback of these devices;
without it, our detectors would enter electrothermal oscillations throughout much of the 
TES transition.

We voltage bias the TESs into the transitions typically at half the normal resistance $R_n$ ({\it i.e.} $R_{\mathrm{bias}} \simeq R_n/2$),
 using $3~\mathrm{m}\Omega$ parallel shunt resistors.  We detect changes in the TES current using a
 SQUID-based time-domain multiplexing
architecture, where all detectors in a set of 32 are sequentially read through a common set
of lines (\cite{deKorte_TDM} and \cite{MUX_chips}).  For \biceptwo, our revisit rate of a given detector is $\sim 25~\mathrm{kHz}$,
ensuring that we Nyquist sample the time-stream up to the 6~kHz roll-off imposed by a
$1.35~\mu\mathrm{H}$ series inductor.

Adapting this technology to the low photon loadings of \spider's long-duration balloon payload poses challenges in detector and instrument design.  In order to take full advantage of the low photon noise levels available at 36~km altitudes, care must be taken to ensure that the bolometers achieve low noise levels and the surrounding instrument contributes minimal additional photon loading.  

\spider\ employs several modifications to the \biceptwo\ detector design to
reduce detector noise.  A meandered leg design reduces conductance to $G\sim15$~pW/K, reducing phonon noise while remaining
within a similar footprint.  This chosen $G$ is not limited by the anticipated loading of $\sim4.5~\mathrm{K}_\mathrm{CMB}$ but rather to have margin on the 300K background in the laboratory while biased on the Al TES.  Lower $G$ leads to naturally slower
detectors, reducing the need to add heat capacity to the island with thick gold (Au)
to maintain stability.  The long cabling in \spider\ prevents us from multiplexing
faster than $\sim20$~kHz, so we control aliasing with larger inductors.

\section{Fabrication}
\label{sec:fab}

We fabricate detectors in monolithic batches of 64 matched pairs at the Microdevices Laboratory (MDL) at the Jet Propulsion Laboratory (JPL).  We fabricate arrays on $350$--$800~\mu\mathrm{m}$ thick silicon substrates chosen for optimal optical coupling.  We deposit $0.7$--$1.2~\mu\mathrm{m}$ low-stress silicon nitride films (LSN) and tune the internal stress to be as low as 150~MPa to ensure mechanical integrity.

\begin{table}[ht]
\caption{Summary of Processing Steps$^*$}
\centering
\begin{tabular}{l l l}
\hline\hline
Material/Function & Deposition &Etchant\\ [0.5ex] 
\hline
1. Low-stress nitride &High temp LPCVD & CHF$_3$ ICP-RIE\\
2. Aluminum TES & e-beam evaporation & Liftoff\\
3. SiO$_2$ Pro-1 & RF sputtered & ICP, CHF$_3$/O$_2$ etch\\
4. Titanium TES & sputtered & ICP, Freon-1202\\ 
5. SiO$_2$ Pro-2 & RF sputtered & ICP, CHF$_3$/O$_2$ etch\\
5. Niobium Ground Plane & Magnetron sputtered & Liftoff\\
6. SiO$_2$ ILD & RF sputtered & ICP CHF$_3$/O$_2$ etch\\
7. Gold resistor & e-beam evaporated & Liftoff\\
8. Niobium Microstrip & sputtered & ICP, Freon-1202\\
9. Gold heat capacity & e-beam evaporated & Liftoff\\
10. Silicon release & substrate & STS DRIE/ XeF$_2$\\
\hline
*Acronyms defined in text & &\\
\end{tabular}
\label{table:steps}
\end{table}

\begin{figure}[ht]
\centering
\subfigure{
\includegraphics[width=1\columnwidth]{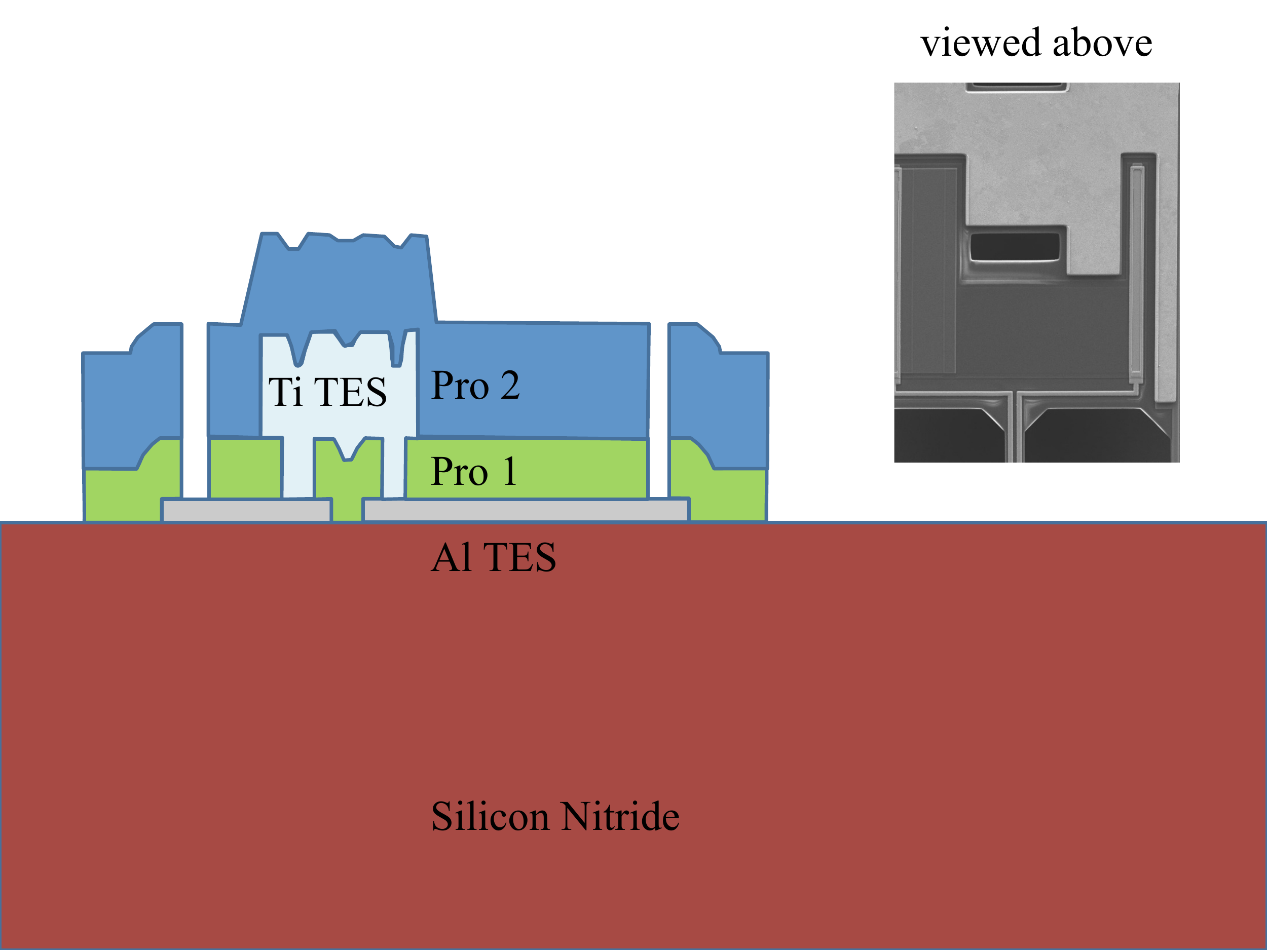}
}
\subfigure{
  \includegraphics[width=1\columnwidth]{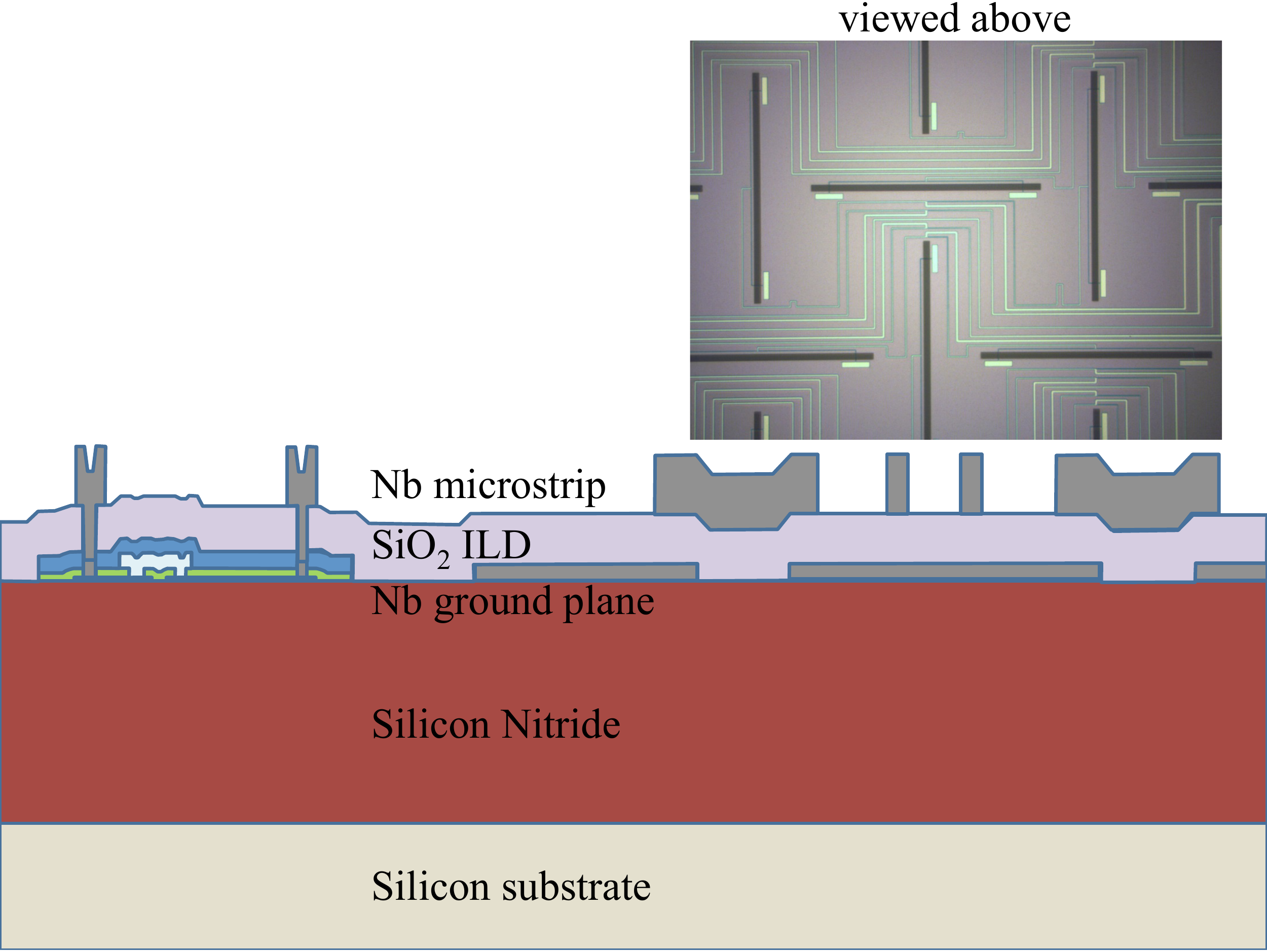}
}
\subfigure{
  \includegraphics[width=1\columnwidth]{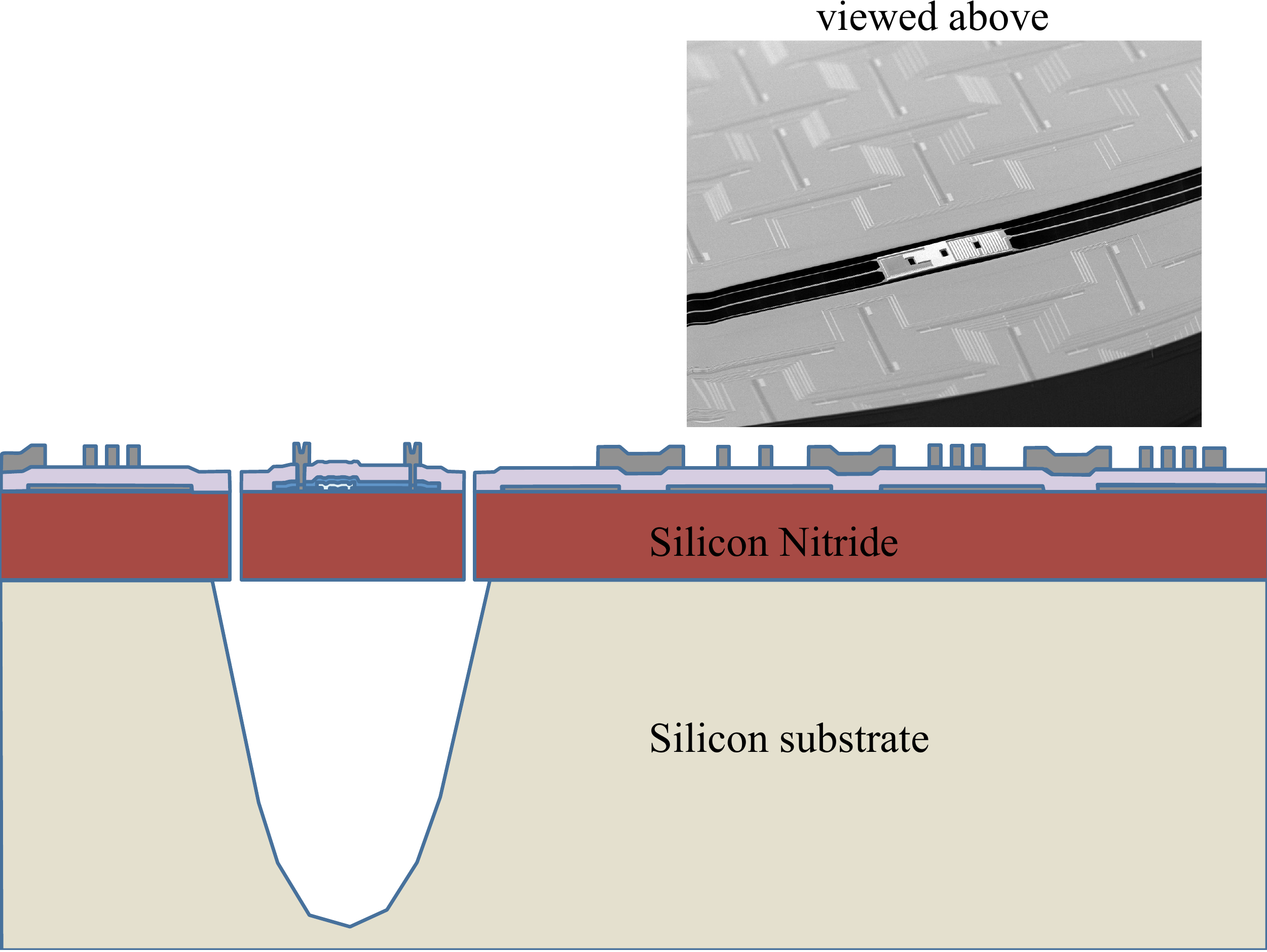}
}
\caption{Cross-section of films in order of fabrication  The aspect ratio between radial and normal dimensions are distorted for clarity.  We include photos of how the device looks face-on for reference.  {\it Upper}: Deposition and etching of films for the TESs and their protect layers.  {\it Middle}: Deposition and etching of antenna and microstrip features.  {\it Lower} Release of TES bolometer.\label{fig:fabrication_stratification}}
\end{figure}

The TES normal state resistance $R_n$ and transition temperature $T_c$ are
sensitive to thickness variability and chemical contamination, so we deposit
and pattern our TESs before the millimeter wave circuitry on a flat and
chemically clean surface.  We first e-beam evaporate aluminum (Al) and pattern the $T_c=1.2~\mathrm{K}$ TES.  We
DC-sputter titanium (Ti) and use inductively coupled plasma (ICP) etching to etch the $T_c=0.52~\mathrm{K}$ TES for
on-sky observing.  Both films are immediately chemically passivated
with RF-sputtered silicon dioxide ($\mathrm{SiO}_\mathrm{2}$) with ICP etched via
holes to allow DC electrical contacts with subsequent layers.  The RF bias ensures that the passivation patches have rounded edge walls so that subsequent films
can make contact over steps.  The top panel of Figure~\ref{fig:fabrication_stratification} illustrates these first steps.
The Al and Ti films make a series DC connection and are later connected to Nb bias lines.

The millimeter wave circuits are fabricated from four films: a niobium
ground plane, a $\mathrm{SiO}_\mathrm{2}$ interlayer dielectric
(ILD), an upper microstrip Nb conductor, and thin Au resistive
termination.  We DC-sputter the Nb films and pattern the ground plane
with lift-off, a technique where we first deposit photoresist and then
dissolve it from under the metal film to be removed.  In principle, 
this film could be etched, but we suspect that the Ti $T_c$ can be altered by this 
step despite the intervening protect layer.  The Nb ground
film defines the antenna slots, bandpass filter inductors, holes
for bolometer release, and safety holes under the bond-pads.  This film also
fills the bias vias down to the TESs.  Lift-off provides rounded side-walls
and thus ensures step-coverage of subsequent films.  The Nb ground-plane covers
more than 90\% of the 100~mm wafer, serving as both the DC and RF ground 
for the detector array.

We have explored several ILD materials, although all currently deployed devices
use RF-sputtered silicon dioxide.  Thickness uniformity of this film is crucial
for detector uniformity across the array.  A 6 inch SiO$_2$ RF sputtering
target and substrate rotation help achieve thickness uniformity to better than
7\% across the array.  We use an inductively coupled plasma (ICP) reactive ion etch
system to etch release holes around the bolometers and vias that allow for a DC
connection to the buried TES structures in the subsequent Nb metallization fabrication step.  

We form the resistive terminations on the bolometer islands with e-beam deposition
and lift-off patterning.  Finally, we DC-sputter the upper Nb microstrip conductor and
etch from that film the microstrip feed network, band-defining filter capacitors, and DC bolometer bias lines and bond-pads using a freon etch.  We have found that lift-off
techniques (used for \biceptwo)  and chlorine-based (BCl$_3$) etches can contaminate the Nb
films at this step, resulting in unacceptable millimeter wave losses in our circuitry as discussed in \S\ref{sec:film_properties} and \S\ref{sec:contamination}.
The middle panel of Figure~\ref{fig:fabrication_stratification} illustrates these intermediate microwave steps that define ground plane, ILD, and microstrip traces.

The final fabrication steps release the bolometer islands, illustrated in
The bottom panel of Figure~\ref{fig:fabrication_stratification}.  We etch through the LSN with
$\mathrm{CHF}_\mathrm{3}$ in an inductively coupled plasma reactive ion etch ICP-RIE system, exposing bare silicon
under the release holes.  We e-beam evaporate and lift-off thick gold
onto the islands to add heat capacity and control the required readout
bandwidth.  We use an STS deep reactive ion system to etch from the front side
and remove silicon from the release holes.  This Bosch etching process uses
a combination of etching and passivation steps to cut vertically through
the silicon wafer, after which we undercut the bolometers with a XeF$_2$ release \citep{Turner:01}.
XeF$_2$ attacks the remaining exposed silicon isotropically, so the small hole
pattern in the photoresist is pre-determined to release only the islands and
legs and to not undercut the remaining antenna structures. Finally, we use
the STS etcher to cut out the square tile with holes for alignment
pins.  Table~\ref{table:steps} summarizes these fabrication steps.

\section{Material Properties for the millimeter wave circuits}
\label{sec:film_properties}

We separately characterize the material properties used in the design of the antenna feed and filter,
in particular the Nb penetration depth and the ILD dielectric constants. These
material parameters are poorly characterized in the literature for millimeter
waves, and they can depend on details of the processing.  We also need to characterize
loss properties and monitor stability over time.  To address these concerns, we
have fabricated companion tiles of test devices where each detector pair receives power
through a single polarization broadband integrated antenna.  The microstrip
feed evenly divides power in a microstrip tee-junction between a device under test (DUT) and a reference bolometer to divide away optical effects from the antenna and fore optics.  We measure the spectral response of the bolometers through Fourier transform spectroscopy (FTS) with a Martin-Puplett interferometer \citep{Martin1970105}, combining the detector time-streams and encoder readings from the moveable mirror's translation stage to form interferograms.  We low-pass filter, zero-path difference, and Hanning apodize the interferogram before Fourier transforming the interferogram into the frequency response $S(\nu)$.

\begin{figure}[ht]
\centering
\subfigure{
\includegraphics[width=1\columnwidth]{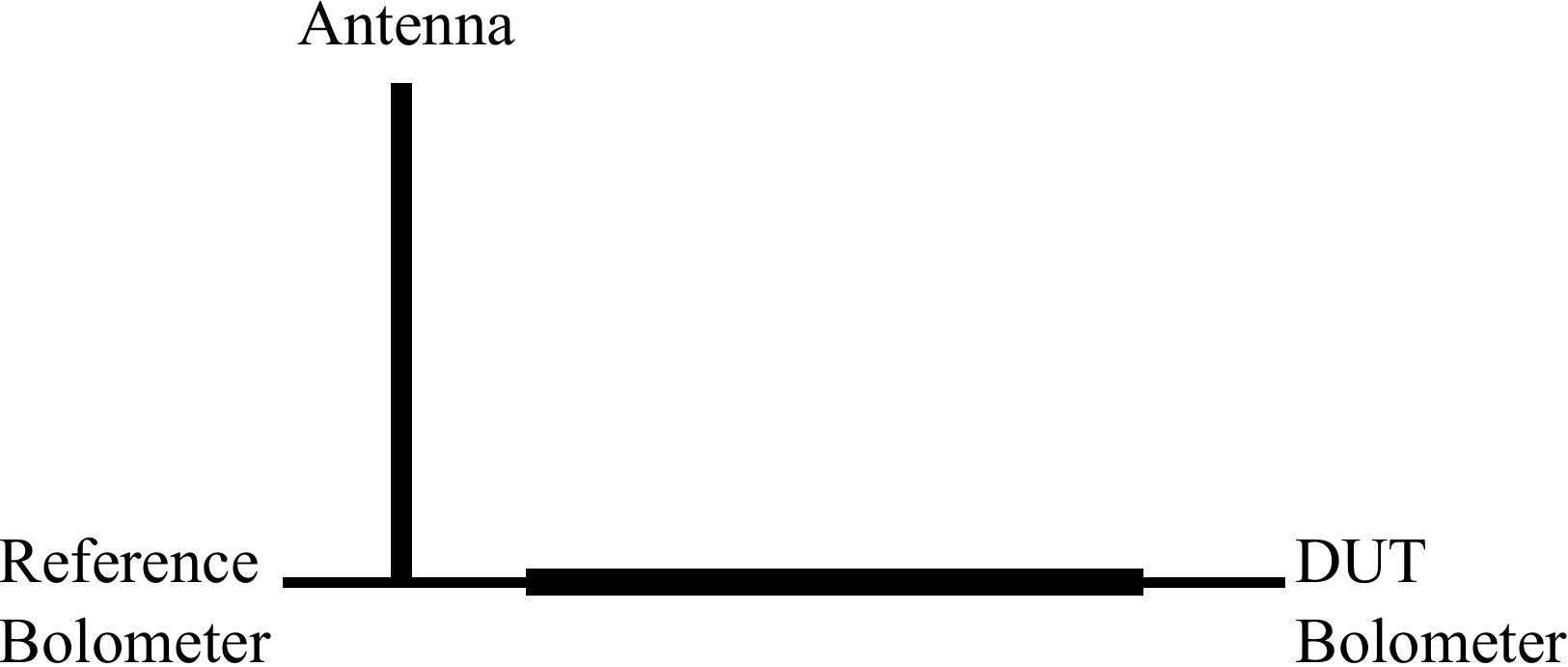}
}
\subfigure{
  \includegraphics[width=1\columnwidth]{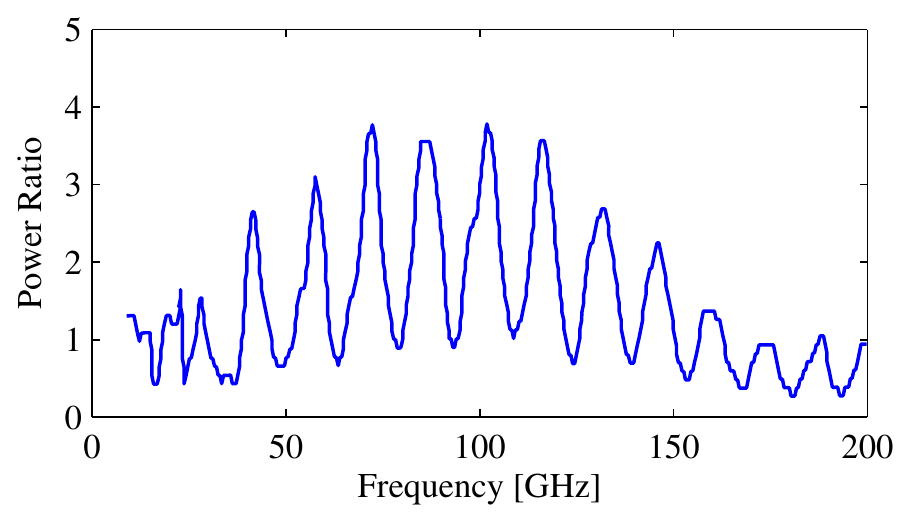}
}
\caption{{\it Upper}: Schematic of a test device for microstrip wave speed, where the mismatched microstrip segment at right forms a Fabrey-Perot cavity.  {\it Lower}: Standing wave pattern in a device's FTS spectrum.  The period is proportional to the transmission line wavespeed.\label{fig:wavespeed_test}}
\end{figure}

We use one device for microstrip wave speed measurements.  The DUT is a microstrip Fabrey-Perot cavity with an impedance intentionally mismatched from the surrounding lines.  The standing wave pattern in the spectrum of the bolometer behind the cavity can be used to determine wave speed.  (See Figure~\ref{fig:wavespeed_test} for a schematic and data sample.)  The wave speed is a function of both the ILD dielectric constant $\epsilon_r$ and the Nb penetration depth $\Lambda$.  Alternatively, the band-defining filter resonance is determined by series inductance dominated by the CPW magnetic inductance.  Data from the band locations and test devices allow us to solve for both $\epsilon_r$ and $\Lambda$ and we summarize these numbers is Table~\ref{table:material_values} for different materials.

\begin{table}[ht]
\caption{Material Properties at millimeter waves}
\centering
\begin{tabular}{c c}
\hline\hline
Parameter & Value \\ [0.5ex] 
\hline
SiO$_2$ $\epsilon_r$ & 3.9\\
Si$_x$N$_y$ $\epsilon_r$ & 7.0\\
Nb $\Lambda$ & 0.1~$\mu$m \\ [1ex]
\hline
\end{tabular}
\label{table:material_values}
\end{table}

Another test device measures loss per length in our transmission lines, where
the device under test is a stretch of line that is several wavelengths long.
We determine loss as a function of frequency by forming the ratio
of FTS spectra seen through the long line and the reference bolometer.  While the
loss tangent has been shown to be dominated by loss in dielectrics, and can be
quite low in some dielectric systems, our measurements determine loss in the microstrip
as fabricated.  Thus far we have tested PECVD SiO$_2$, PECVD Si$_x$N$_y$ and evaporated
SiO$_2$ in this manner.

\begin{figure}[ht]
\centering
\subfigure{
\centerline{\includegraphics[width=1\columnwidth]{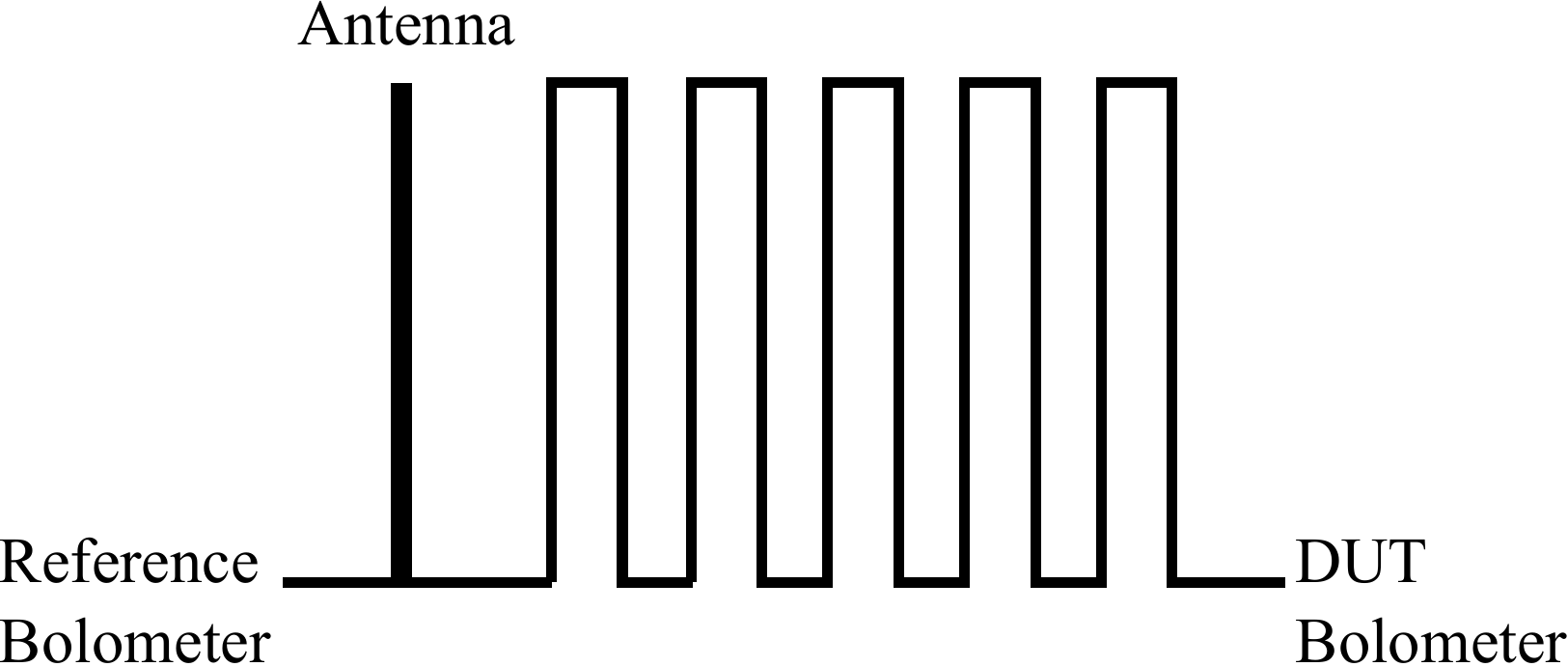}}
}
\subfigure{
 \centerline{\includegraphics{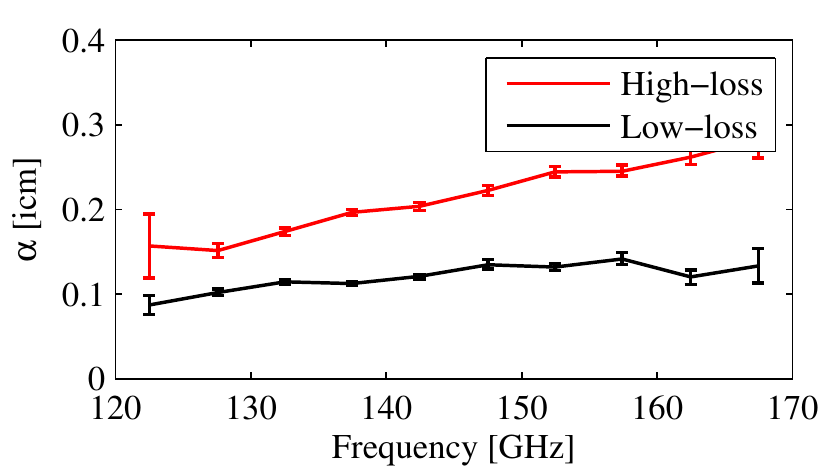}}
}
\caption{{\it Upper}: Schematic of a test device for microstrip loss rates, where the DUT is a stretch of transmission line several wavelengths long.  {\it Lower}: Sample spectra showing power loss per unit length ($\alpha = -d \left(\log P\right)/dx$) in films exhibiting low and high loss.  Nonlinear increase in loss with frequency has been an indicator of bad process parameters in need of adjustment.\label{loss_per_length}}
\end{figure}

This loss test capability became particularly important during a period of time 
when the optical efficiencies of our fabricated devices were unusually poor.  
This test program allowed us to show that the effective loss tangent of our microstrip had a $\nu^2$ frequency  dependence (see Figure~\ref{loss_per_length}) when using all three of the dielectrics above.  This common problem present with three ILDs from different deposition systems suggested that the loss was generated in the Nb films.  Ultimately, these measurements helped us to identify and correct a modest tensile stress in the Nb as deposited by sputtering.  Further, the loss appears to be sensitive to the reactive ion etch chemistry used in the Nb process.  Controlling the film stress at a nominal compressive -100 to -300 MPa and changing the Nb etchant from BCl$_3$+O$_2$ to Fl$_2$+O$_2$ resulted in lower loss tangent with a $\nu^1$ dependence (see Figure~\ref{loss_per_length}).

\section{Beam synthesis}
\label{sec:beams}

\begin{figure}[htbp]
\centerline{\includegraphics[width=1\columnwidth]{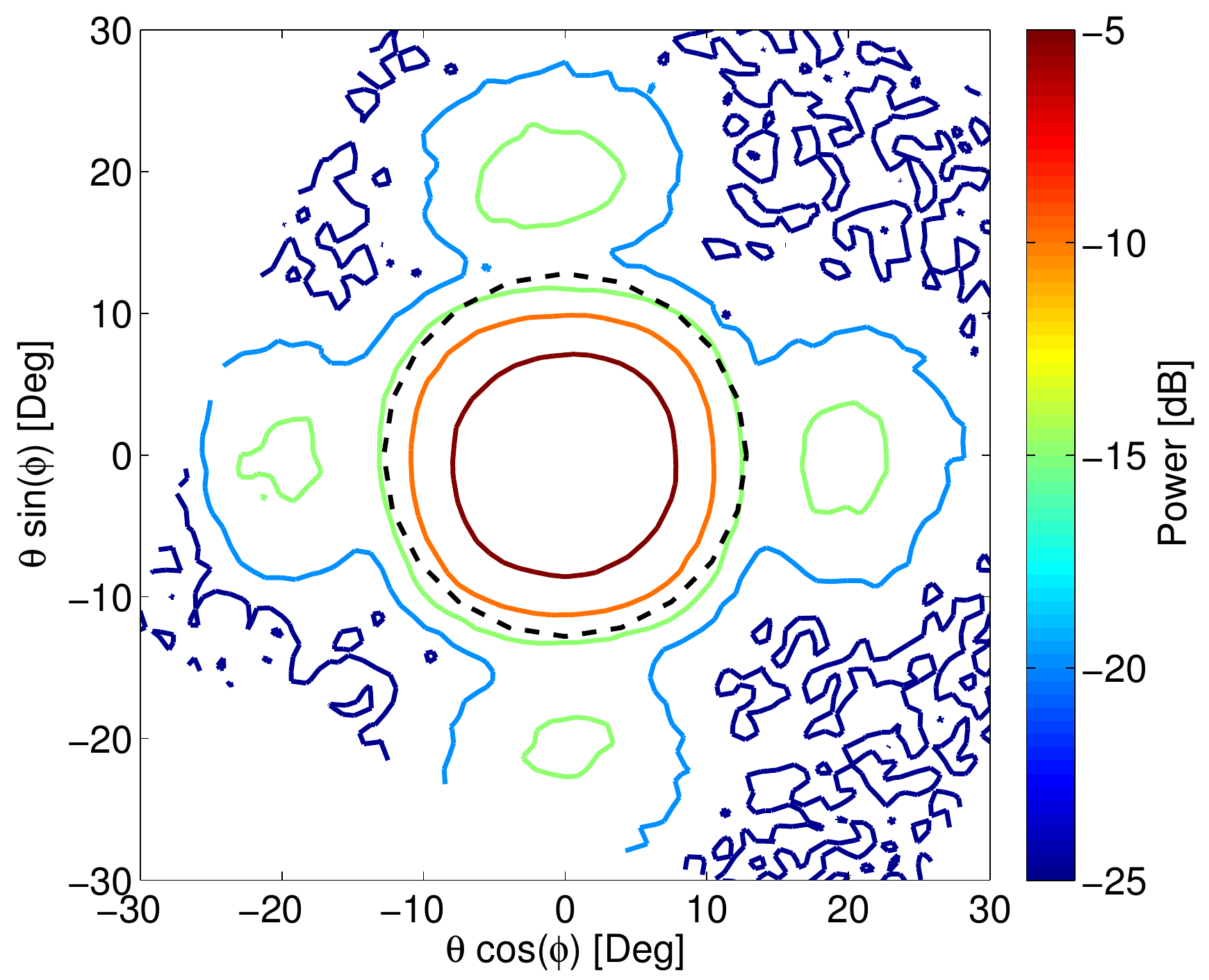}}
\caption{\label{fig:far_field_pattern}Sample far-field detector pattern measured in a test cryostat without an optical stop.  Power is normalized to peak on boresight and the dashed line indicates where the f/2.2 camera stop would lie.}
\end{figure}

The antenna array combines waves from the sub-antennas with equal amplitude and phase to synthesize a uniform illumination, which generates a sinc pattern in the antenna far field.  Using a test cryostat, we characterized the antenna far field pattern and confirmed that the antenna has a $1/e$ beam waist of 4.1$^\mathrm{o}$ ($\mathrm{FWHM}\sim 14^\mathrm{o}$) and side-lobe level of -12~dB relative to peak response, as expected from the design.  A sample far-field pattern is shown in Figure~\ref{fig:far_field_pattern}, taken in a cryostat without lenses or a stop.  For a detailed beams characterization in the complete \biceptwo\ and \keck\ cameras, we refer the reader to the \biceptwo\ and \keck\ Beams Paper \citep{b2beams15}.

Our ground-based experiments difference detector pairs at the time stream level
to suppress unpolarized common-mode noise from the atmosphere.  However, differencing
at this early point in the analysis pipeline can allow temperature anisotropies to leak
into polarization if the detector beam patterns are mismatched.
This is an especially acute challenge to small aperture experiments because the beam
size couples to large temperature gradients at degree scales.

Our team has developed an analysis technique, called deprojection, that discards contaminated modes.  As described in
the \biceptwo\ Systematics Paper \citep{b2syst15}, the pipeline can remove temperature leakage through relative-gain mismatch (monopole moment of beams), first derivatives of temperature through displaced centroids (dipole moment of beams), and second derivatives of temperature
through differential beam widths and ellipticities (quadrupolar moments of beams).

The detector design is most prone to centroid displacement, as characterized in
the Beams Paper \citep{b2beams15}.  We have identified two mechanisms responsible for beam displacement
and implemented design and fabrication fixes to suppress their contamination below the
$r=0.1$ level \citep{obrient12}.  These fixed detectors are deployed in \keck\ and \spider\ and Figure~\ref{fig:measured_dipoles} shows the centroid steer before and after these corrective measures.

\subsection{Parasitic microstrip cross-talk}

\begin{figure*}[ht]
\centering
\subfigure{\includegraphics[width=1\columnwidth]{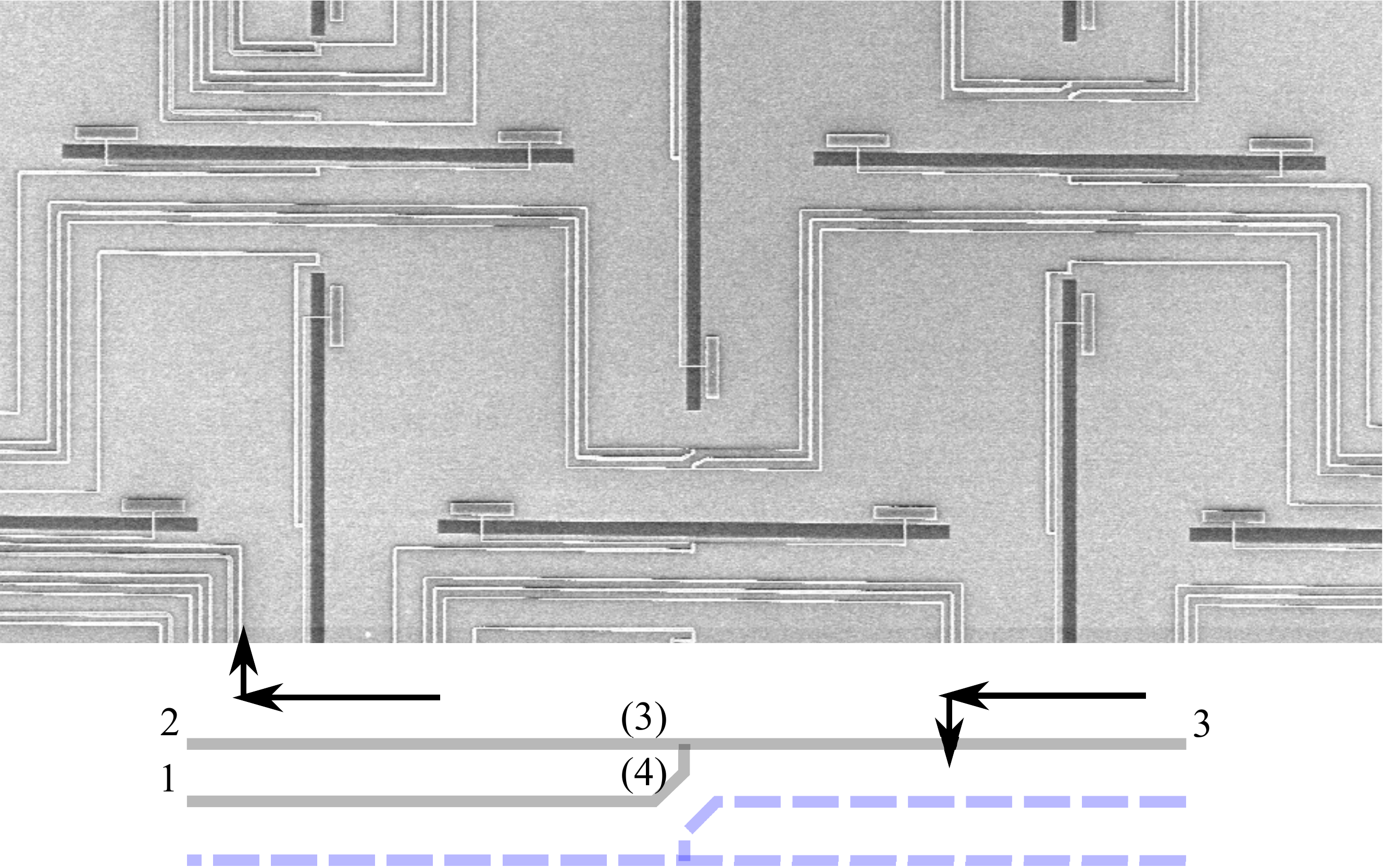}}
~~
\subfigure
{\hspace{0.7em}\includegraphics[width=1\columnwidth]{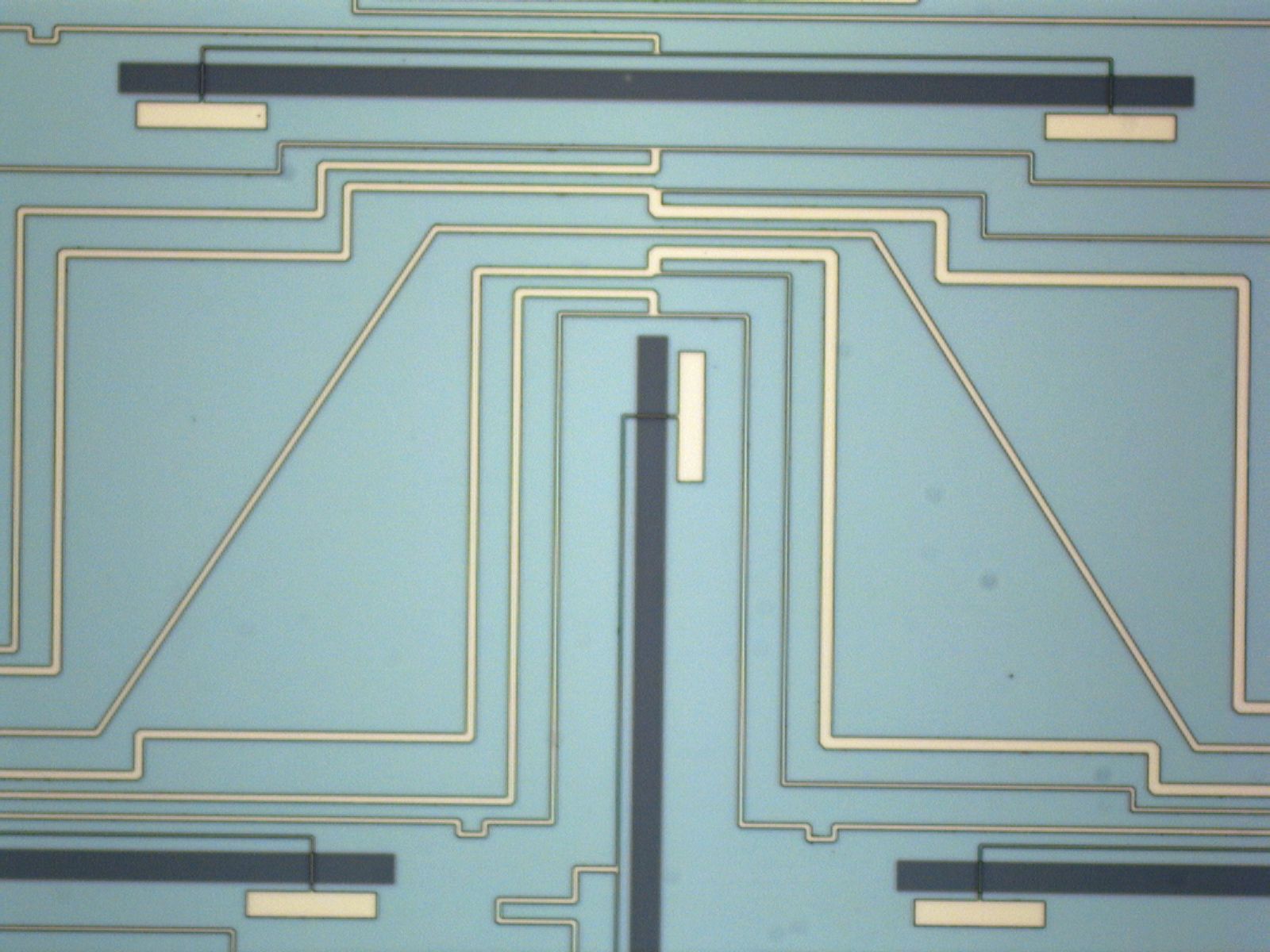}}
\caption{Details of microstrip layout in the center of the antenna feed.  {\it Left}: Center of the \biceptwo-era design.  The Cartoon under the picture shows the effective coupler circuit and the ports corresponding to the scattering parameters in Equation~\ref{eqn:coupler}.  The arrows are phasors that illustrate how the intended waves (horizontal) combine with the parasitic coupled waves (vertical).  {\it Right}: Modern design with reduced coupling due to greater spaced lines.  The extra lengths of line that form compensating phase lags are visible (e.g. bottom of picture), although these are not necessary with the suppressed microstrip coupling.\label{fig:spaced_lines}}
\end{figure*}

The microstrip lines in the feed networks' horizontal arms must fit between the slot
sub-radiators, crossing them only at the intended feed points.  To achieve this, the lines
must be in close proximity, and for the 150~GHz \biceptwo\ devices, many lines are
separated by only $\sim$10 dielectric thicknesses.  This separation would be adequate to avoid
cross-talk for short stretches of lines, but some of the line pairs run parallel for over four
wavelengths.  The longest of these pairs are those running from the sides to the center to
begin the horizontal tree arms, coupling power to one of the two lines they intentionally
split power between, as seen in Figure~\ref{Summing_tree_topology} and summarized in the bottom of the left panel of Figure~\ref{fig:spaced_lines}.

In a reverse-time picture, power from port 1 (defined in Figure~\ref{fig:spaced_lines}) should be evenly divided between ports 2 and 3.  The currents in a pair of parallel lines can be expressed as a superposition of even and odd modes, where the even mode has the same current magnitude and directions in each line and the odd mode has the same magnitude but opposite directions.  Coupling between the two parallel lines, however, results in the two modes having different field concentrations in the air and dielectric volumes, and so induces a difference in wave speeds between the modes.  Waves passing from port 1 to internal port (4) thus also couple waves into internal port (3) that lag by 90$^\circ$ in phase:  
\begin{eqnarray}
 S_{(4)1}&=&(S^e_{(4)1}+S^o_{(4)1})/2=(e^{-jk_e\ell}+e^{-jk_o\ell})/2 \nonumber\\
 &=& \quad e^{-j(k_e+k_o) \ell/2}\cos ((k_e-k_o)\ell/2)      \nonumber \\
 S_{(3)1}&=&(S^e_{(3)1}+S^o_{(3)1})/2=(e^{-jk_e\ell}-e^{-jk_o\ell})/2  \nonumber \\
 &=& -je^{-j(k_e+k_o) \ell/2}\sin ((k_e-k_o)\ell/2)
 \label{eqn:coupler}
\end{eqnarray}
where $k_e$ and $k_o$ are the even and odd mode wavenumbers and $\ell$ is the
parallel coupling length.  This circuit acts as an unintentional reverse-wave
coupler, where impedance mismatches between even and odd modes are small, but
wavespeeds are not \citep{1124303}.  The waves entering the microstrip tee-junction from
internal port (4) add in quadrature to the intended waves from internal port (3),
advancing the total phase on the side opposite the vertical tree and retarding it
on the other.  This phase-step steers the beams off boresight away from the vertical
summing trees and is where the left-right symmetry splitting occurs in the feed.
This effect produces the repeatable horizontal centroid steer seen in the cameras' near
field.  We have reproduced this in simulations using HFSS that account for the  upper conductor finite thickness.

As seen in the right panel of Figure~\ref{fig:spaced_lines}, we have re-designed the antenna feeds for the \keck\ and \spider\ detectors to greatly increase spacing between the lines and therefore reduce the parasitic coupling effect.  We also include adjustable phase-lag lines before each slot to remove residual phase-error and to synthesize matched beams, although the current antenna-feeds' increased spacing renders the corrective phases unnecessary.  These phase-lags are visible in the bottom of the right picture in Figure~\ref{fig:spaced_lines}.

\subsection{Niobium contamination}
\label{sec:contamination}
Magnetic fields can penetrate into a superconductor by a characteristic
depth $\lambda_{eff}$.  Impurities in niobium films scatter Cooper pairs
with a mean free path $\ell$, increasing the penetration depth beyond
the London depth $\lambda_L \sim 50~\mathrm{nm}$ of pure niobium to
$\lambda_{eff}=\lambda_L\sqrt{\xi_o/\ell}$, where $\xi_o \sim 40~\mathrm{nm}$
is the Cooper pairs coherence length.  As displayed in Table~\ref{table:material_values},
our films are typically measured to have $\lambda_{eff}\sim100~\mathrm{nm}$,
suggesting mean-free path of $\ell=10~\mathrm{nm}$, well within the
``dirty'' limit where film cleanliness can impact circuit performance.

Nonuniform contamination can produce nonuniform kinetic inductance, which can spatially perturb the wavespeeds in the microstrip summing tree.  Variations in wavespeed can steer beams off boresight.  Additionally, the summing tree does not treat the two polarizations identically, and as a result, they can be steered differentially.  We have observed that the tiles with the largest scatter in beam centroid position correspond to those with the largest vertical dipole components.  We also expect the tree to induce larger steering in the vertical than horizontal because slots along rows combine immediately in the horizontal tree, resulting in less integrated phase error than those along columns that combine in the vertical tree after horizontal summing.

We have built models for how our antenna-feeds perform with film
gradients.  We subdivide the circuit into short sections of transmission
lines and tee-junctions and use each section's location in the detector to assign unique film properties.  We construct simple scattering matrices
for each section and cascade them into one large $289\times 289$ matrix per
polarization\citep{1083357}.  From this we can compute slot illumination
patterns and thus far-field patterns.  We find that 20\%
variations in $\lambda_{eft}$ across the array can produce differential pointing that is 10\%
of the beam FWHM in the vertical directions, and half that on the
horizontal, similarly matching our observed scattering in pointing.

\begin{figure*}[ht]
\centering
\subfigure{
\includegraphics[width=1\columnwidth]{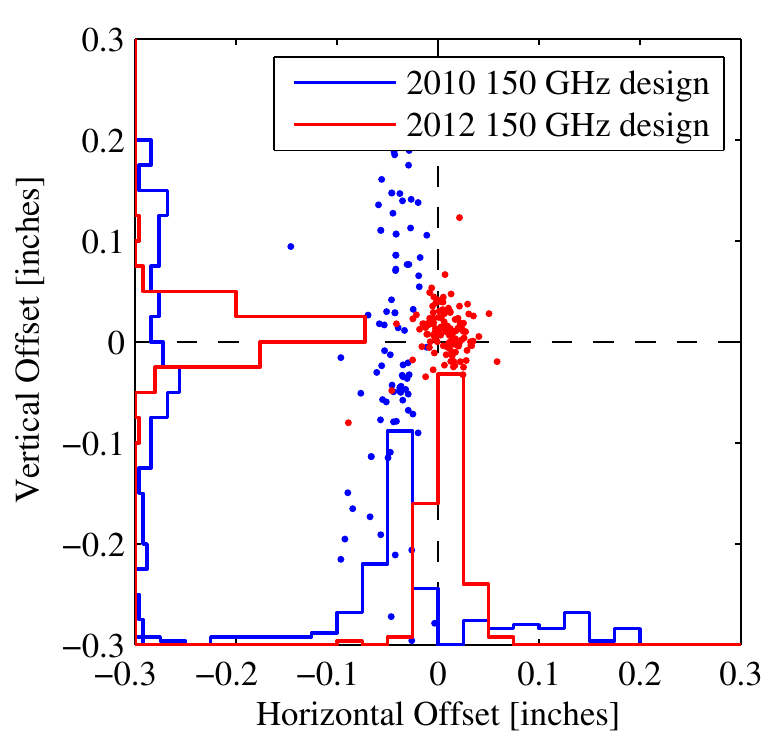}
}
~~
\subfigure{
  \includegraphics[width=0.95\columnwidth]{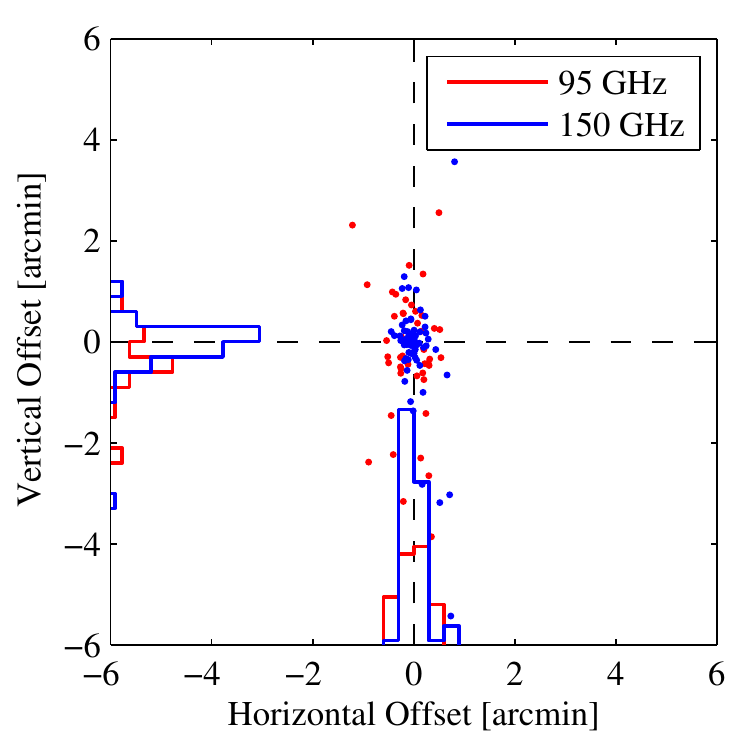}
}
\caption{{\it Left}: Beam centroid displacements in the near field before and after fixes were made to the etch recipe.  The performance shown in left panel would allow a \bicepone-style pipeline without deprojection to constrain $r$ to 0.1 \citep{Takahashi_B1}.  The deprojection pipeline described in \citep{b2syst15} allows $r$ to be constrained to yet lower levels.  {\it Right}:  Beam centroid displacements in the far field for two different colors after the recipe was fixed.  \label{fig:measured_dipoles}}
\end{figure*}

We defined our microstrip lines in early tiles with the lift-off technique that we use for the Nb ground plane, as described in \S\ref{sec:fab}.  Several devices have shown discoloration in this step, leading us to speculate that the Nb leaches organic materials from the resist during lift-off.  These observations and modeling inspired a switch to an etch-based means of defining the Nb microstrip lines (described in \S\ref{sec:fab}).  This simple fix reduced the scatter in centroid location to $\sim1$\% of Gaussian width $\sigma$, and the right panel of Figure~\ref{fig:measured_dipoles} shows the centroid alignment between polarization pairs for both colors of deployed focal planes.

\subsection{Direct stimulation of bolometers}
\label{sec:direct_stim}
Optical power is meant to reach the bolometers only through the antenna and microstrip feed network.  However, photons can directly excite responses in the detectors which is of particular concern in a design where there are no horn blocks to shield the sensors themselves.  
In \biceptwo\ pre-deployment testing of early generations of detectors, response to out-of-band
power was detected at levels 3\%-4\% of the total response, with near field and far field
angular response patterns consistent with direct stimulation of the bolometer islands.  Simulations in HFSS and CST suggest that the ground plane on the bolometer islands can be inductively held at a different voltage than the surrounding ground plane through the microstrip ground that we deposited only on one leg in early generation prototypes.  This voltage can drive millimeter-wave currents through the gold-termination, resulting in a direct stimulation of the detectors not through intended antenna and microstrip feed.  Steps taken prior to \biceptwo\ deployment to minimize this coupling included the addition
of metal mesh low-pass edge filter above the focal plane and several design changes
to the bolometer islands themselves.  The island leg design was modified to narrow the
width of the opening in the ground plane surrounding the island, and ground plane
continuity was extended onto the island by metalization of the four outer support legs.

With these modifications in place, we have taken  \biceptwo\ and \keck\ 150~GHz camera optical efficiency measurements with and without high-pass ``thick-grill'' filters that obstruct power below 200~GHz; after accounting for the filter's filling factor, we found that the optical response through the filter was $\sim0.5$\% of that without filtering.  This small leakage suggests limited ``blue-leaking'' that would allow above-band power to excite response in the bolometer.  We have also measured the response of our detectors to a chopped thermal source on boresight through a polarizing grid at different angles; we have found that the crossed response is similarly $\sim0.5$\% of the co-polarized response, consistent with known multiplexer cross-talk levels.  If the detectors are acting as a direct absorber, then their small area compared to the antenna's should provide a broad angular response.  As a result, we expect that the $f/2.2$ stop in the cameras for these experiments helps limit the direct stimulation and we have found that tests of devices in cameras with faster optics can have higher direct stimulation levels.

\section{Array properties and uniformity}
\label{sec:array_prop}
Our high attained sensitivities are a result of our high detector yield, a figure that depends on attaining uniform detector properties across the array.  This uniformity also helps to mitigate some potential sources of systematic error.  This section describes array uniformity of a variety of properties.

\subsection{Bolometer thermal conductance $G$}
Thermal fluctuation noise across the bolometer legs is the largest internal source of noise in \biceptwo, and is given by:
\begin{equation}
\mathrm{NEP_G}^2=4kT_c^2 G_c F(T_c,T_o)
\label{eqn:Bolo_NEP}
\end{equation}
where the ``Mather factor'' $F(T_c,T_o)$ accounts for thermal gradients across the bolometer legs and varies between 0.5 and 1 \citep{Mather:82}.  The saturation power of the detector is given in Equation~\ref{Psat}.  Note that the leg thermal conductance $G$ is a function of temperature.

\begin{figure}[htbp]
\centerline{\includegraphics[width=1\columnwidth]{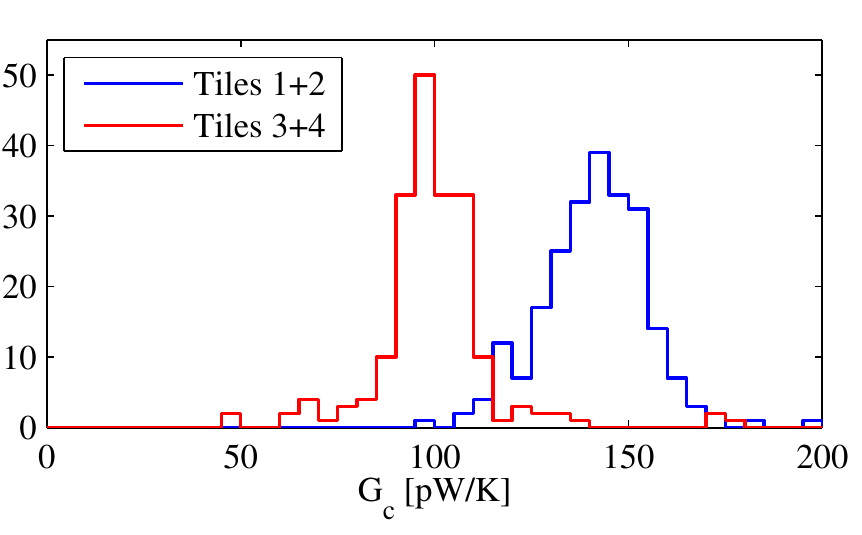}}
\caption{\label{fig:g_hist}Thermal conductance at $T_c$.  This figure demonstrates repeatability within and between tiles.}
\end{figure}

The legs' total parallel thermal conductance $G$ needs to be chosen to keep thermal carrier noise subdominant to photon shot and Bose noise, but avoid saturation under on-sky loading.  We try to have the total saturation power (Joule heating from the bias circuit plus optical loading on the detectors) exceed the optical loading ($\sim4-6$~$\mathrm{pW}$ at 150~GHz in \biceptwo) by a factor of two.  This aggressive safety factor is only possible if we have tight control of our processing parameters that determine $G$ and $T_c$ (discussed in the next subsection).  Our achieved repeatability between detectors in a tile and from tile-to-tile are shown in Figure~\ref{fig:g_hist}.  Our time-division multiplexing readout requires that detectors in a common readout column be commonly voltage biased (\cite{TDM_MUX} and \cite{MUX_chips}), and our control of $G$ within a tile allows this.

The XeF$_2$ release only has a 10:1 selection ratio between silicon nitride and silicon, and variation in bolometer leg thicknesses from nitride etching would seriously compromise these numbers.  For this reason, we oxidize the wafers prior to nitride deposition.  The thin $\sim$100~nm SiO$_2$ film has a much higher 100:1 selection ratio between oxide and silicon and thus protects the legs from XeF$_2$ attack \citep{1257354}.  This oxide film ensures more uniform leg cross-sections across the detector arrays, and thus more uniform bolometer $G$s.

\subsection{Transition temperature $T_c$}

\begin{figure}[htbp]
\centerline{\includegraphics[width=1\columnwidth]{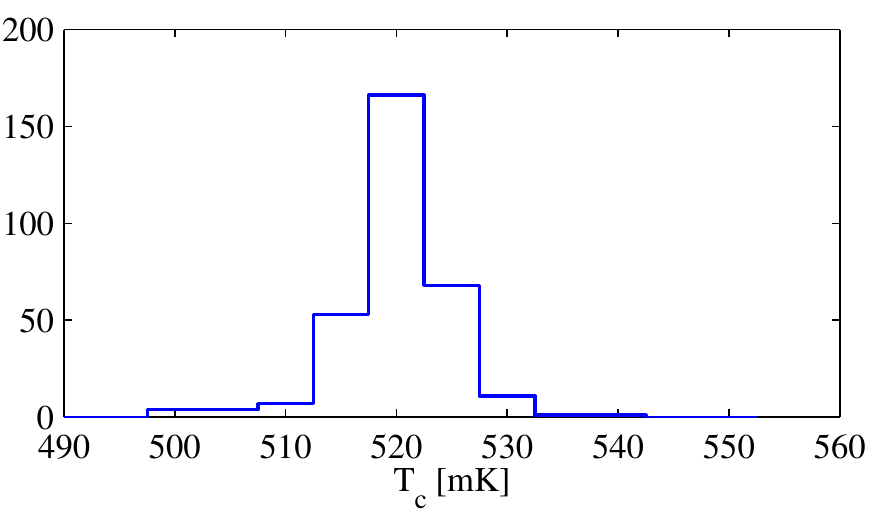}}
\caption{\label{fig:Tc_hist}Titanium TES transition temperature for four tiles in \biceptwo.}
\end{figure}

The balance between the between the internal $^{3}$He/$^{3}$He/$^{4}$He fridge cooling power and thermal loading lets the focal plane cool to $\sim250$~$\mathrm{mK}$.  We use heaters to operate the focal plane at an elevated $280$~$\mathrm{mK}$ so we have margin to control this temperature through active feedback loops while still maintaining photon-noise-limited sensitivity.  This choice provides the detectors with a bath temperature $T_b=280$~$\mathrm{mK}$.  For bolometer legs with n=2.5, $\mathrm{NEP_G}$ is minimized at an island temperature of $420$~$\mathrm{mK}$, although this is a broad-optimum with a gentle slope at higher temperatures.  For on-sky operation, electrothermal feedback locks the bolometer to the titanium TES's $T_c\sim$~$520$~$\mathrm{mK}$.  While this temperature is higher than optimum, it provides some extra margin against our detectors latching into a pure superconductive state while only increasing the NEP from thermal conduction by 4\%.  By selecting a pure superconductor for this TES instead of a bilayer (\cite{5067232}, \cite{Gildemeister_TES} and \cite{2005ApPhL..86k4103M}), we obtain repeatable transitions devoid of the multiple transitions that can result from a bilayer's complicated chemistry.  Deposition on bare wafers and passivation of oxide protection layers also help maintain uniform values between and within tiles.  As seen in Figure~\ref{fig:Tc_hist}, our titanium $T_c$ is uniform across tiles and even between tiles, maintaining uniform saturation powers (Equation~\ref{Psat}) within a multiplexer column, necessary to find a common bias point.

\subsection{Normal resistance $R_n$}

\begin{figure}[htbp]
\centerline{\hspace{-0.7em}\includegraphics[width=1\columnwidth]{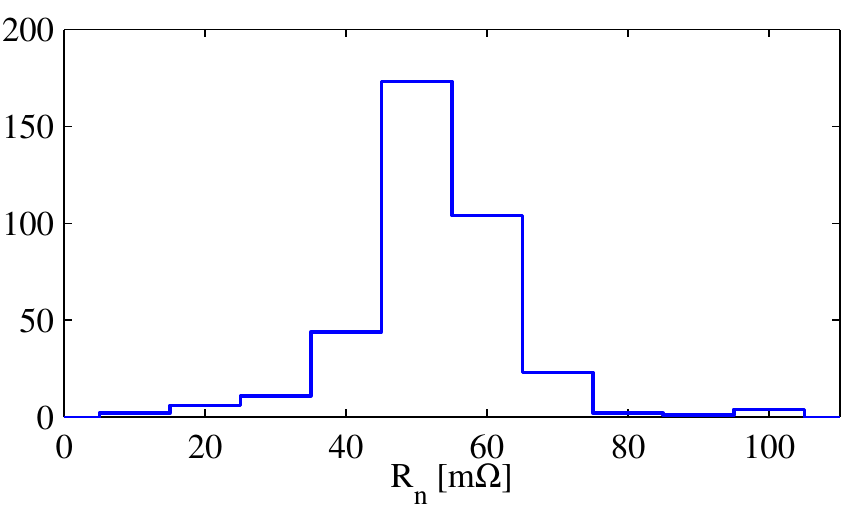}}
\caption{\label{fig:Rn_hist}Titanium Normal Resistance for four \biceptwo\ tiles}
\end{figure}

TES bolometers will only experience electrothermal feedback if voltage biased, and we bias our nominally $\sim60$~$\mathrm{m}\Omega$ Ti TESs with a 3~$\mathrm{m}\Omega$ shunt resistor.  The Al TES is even higher in resistance, so both are far larger than the shunt.  However, this condition is only met if the Ti TES resistance is repeatably this large.  We also need them repeatable in value so the detectors will have a common latch resistance and thus a common bias point where they will experience strong feedback.  Figure~\ref{fig:Rn_hist} shows that these values are indeed uniform between and within tiles.  Just as was true for $T_c$ uniformity, our recipe of a pure TES deposited as first layers on the tile, protected with an oxide film, helps maintain repeatable performance.

\subsection{Time constant $\tau$ and loop gain $\mathscr{L}$}
The effective thermal time constant of an ideal voltage-biased TES bolometer is given by
\begin{equation}
\tau=\frac{G/C}{1+\mathscr{L}(V)},
\label{Bolo_tau}
\end{equation}
where the intrinsic bolometer time-constant $C/G$ is decreased by the effective loop gain of electrothermal feedback, $\mathscr{L}$.  The responsivity of our detectors is the change in measured current for a change in incident optical power, given by
\begin{equation}
s=\frac{-1}{V_b}\frac{\mathscr{L}}{\left( \mathscr{L}+1 \right)}\frac{1}{\left( 1+i\omega\tau \right)}.
\label{Bolo_responsivity}
\end{equation}
Both of these expressions experience corrections for non-idealities such as finite shunt resistance.  Equations~\ref{Bolo_tau} and \ref{Bolo_responsivity} indicate that a high loop gain $\mathscr{L}$ will increase the speed of the sensor and simplify its responsivity ($s \approx -1/V_b$).  Operating the bolometers in this limit also maintains a fast detector response (low $\tau$), giving rise to an approximately flat detector transfer function for the frequencies of interest, which for our ground-based experiments are $f\lesssim2$~$\mathrm{Hz}$~\citep{Irwin_hilton_thy}.

\begin{figure}[ht]
\centering
\subfigure{
\includegraphics[width=1\columnwidth]{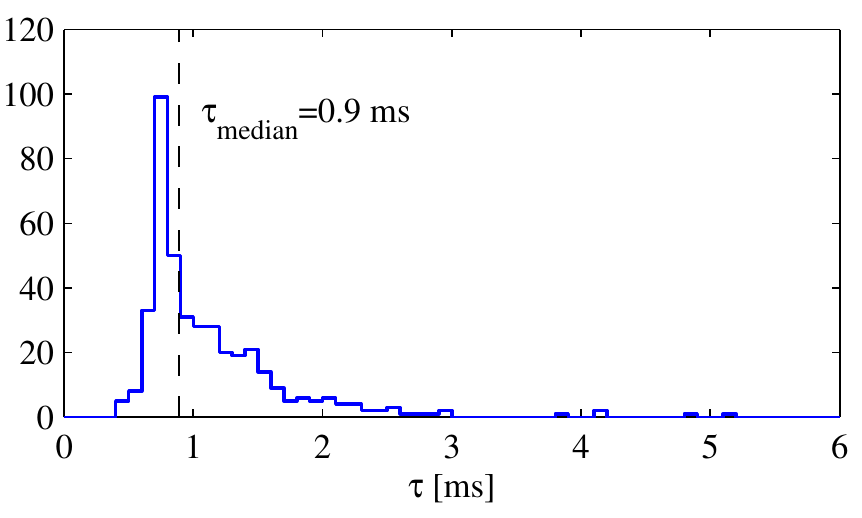}
}
\subfigure{
  \hspace{0.7em}\includegraphics[width=0.98\columnwidth]{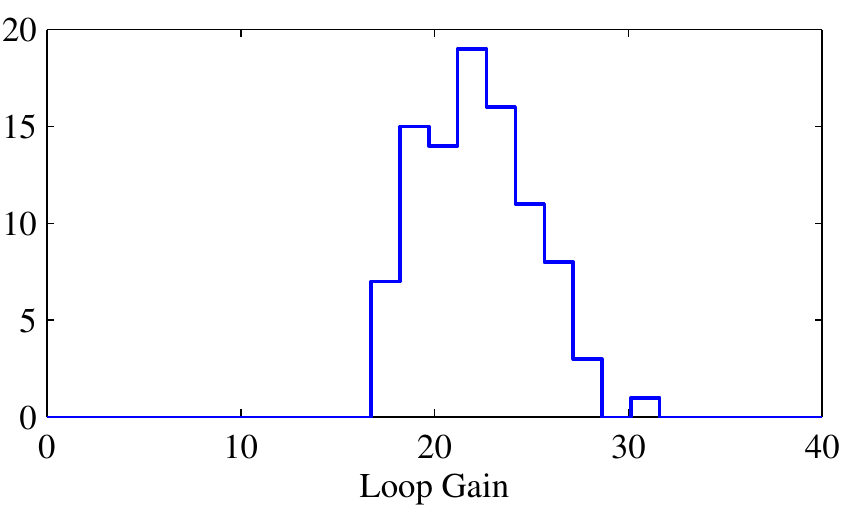}
}
\caption{{\it Upper}: Time constants of \biceptwo.  These were only carefully measured in transition.  {\it Lower}: Loop gain of \spider.  We measured time constant at a variety of biases for \spider\ characterization, allowing a proper measurement of loop gain.  Both sets of detectors had similar $\tau\sim0.9$~$\mathrm{ms}$ biased time-constants.\label{fig:tau_L}}
\end{figure}

In order to maintain stability against electrothermal oscillations, the detectors' thermal bandwidth must not exceed that of the electrical bias circuit.  The \biceptwo\ bias circuit includes a $1.35~\mu\mathrm{H}$ inductor in series with the $\sim3~\mathrm{m}\Omega$ bias resistor to avoid aliasing above a roll-off of $R/L\sim5-6~\mathrm{kHz}$.  To limit the TES bandwidth, we deposit thick $\sim2~\mu\mathrm{m}$ gold onto the \bicep2\ detectors that adds an additional $C\sim 0.5~\mathrm{pJ/K}$ heat capacity to the bolometer island.  The resulting time constants at typical science biases are shown in Figure~\ref{fig:tau_L}.

\spider\ detectors are intrinsically slower due to their lower leg conductance.  It is thus not necessary to add as 
much gold to the island to ensure stability; typically only $\sim$0.5~$\mu$m is deposited.  Time constants
on transition are found to be similar to \biceptwo's.  For \spider\ detectors we have measured time constants
in response to optical square-wave excitations at a range of bias voltages (and thus TES resistances), allowing 
us to infer that the normal time constants (without loop gain) are $G/C\sim30~\mathrm{ms}$ and that 
loop gains are $\mathscr{L}\sim20-30$ at $R=0.6R_n$ (Figure~\ref{fig:tau_L}).  \biceptwo\ TESs 
should have similar loop gains.  We note in passing that fast bolometer time constants are useful for 
limiting the effect of particle radiation ({\em e.g.}, cosmic rays) on detector time streams in balloon- and 
space-borne instruments.

\subsection{Spectral Response}
\begin{figure}[htbp]
\centerline{\includegraphics[width=1\columnwidth]{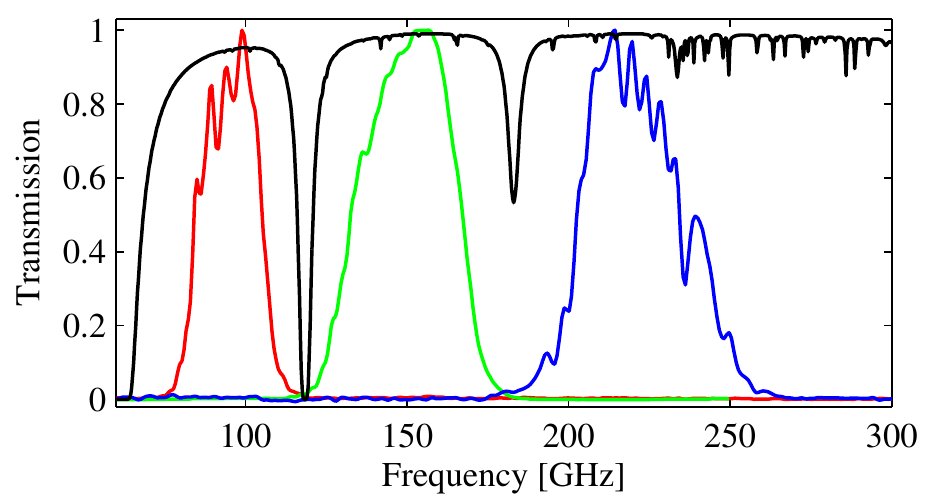}}
\caption{Measured detector spectra for devices designed for 95~GHz (red),
150~GHz (green) and 220~GHz (blue).  The data for the 150~GHz are from \biceptwo,
while the others are from a test cryostat. Winter atmospheric transmission at the South Pole is overlaid
in black.  \label{fig:measured_spectra}}
\end{figure}

Our cameras use a series of low-pass filters, both absorptive plastic and reflective metal-mesh, to limit thermal loading on the focal plane and above-band response of our detectors.  However, we rely upon the integrated microstrip filters described in Section~\ref{sec:filters} to avoid the atmospheric lines immediately adjacent to our observing bands.  Figure~\ref{fig:measured_spectra} shows measured response $S(\nu)$ averaged across a focal plane for three different spectral channels (in three different focal planes), demonstrating that this technology does indeed avoid atmospheric features.

\begin{figure}[ht]
\centering
\subfigure{
\hspace{-0.1em}\includegraphics[width=0.99\columnwidth]{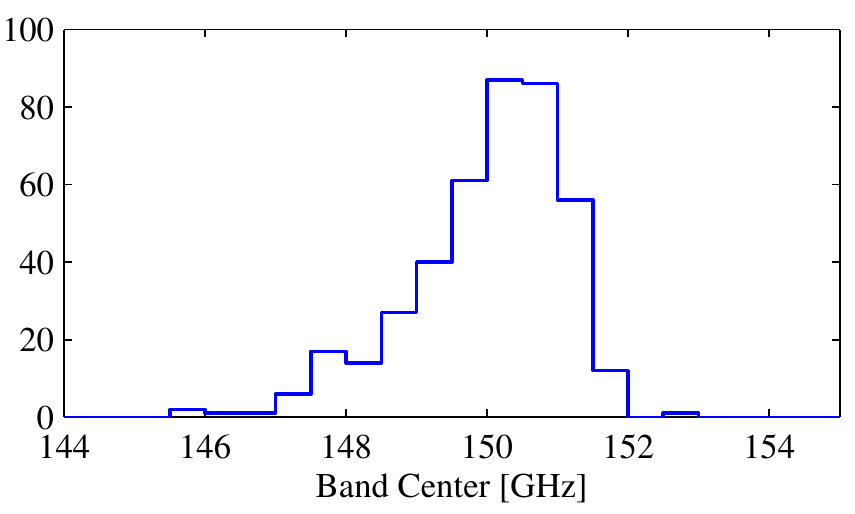}
}
\subfigure{
  \includegraphics[width=1\columnwidth]{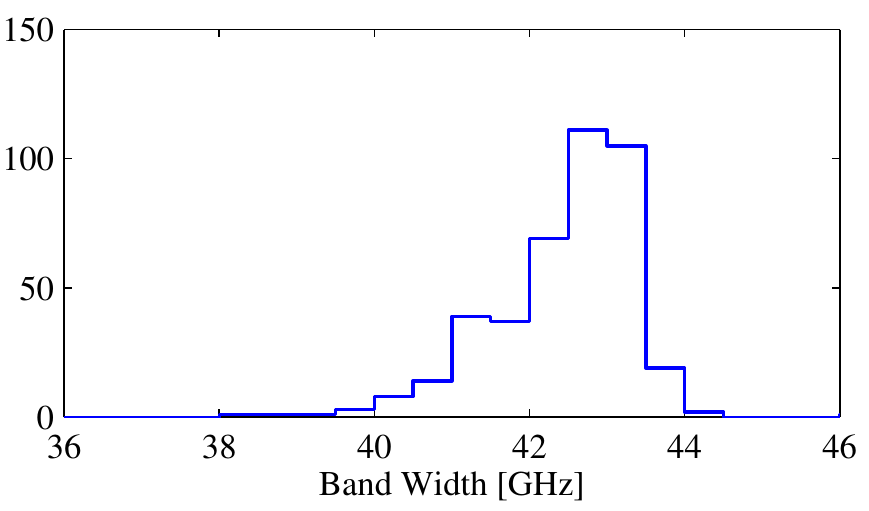}
}
\subfigure{
  \hspace{0.8em}\includegraphics[width=0.98\columnwidth]{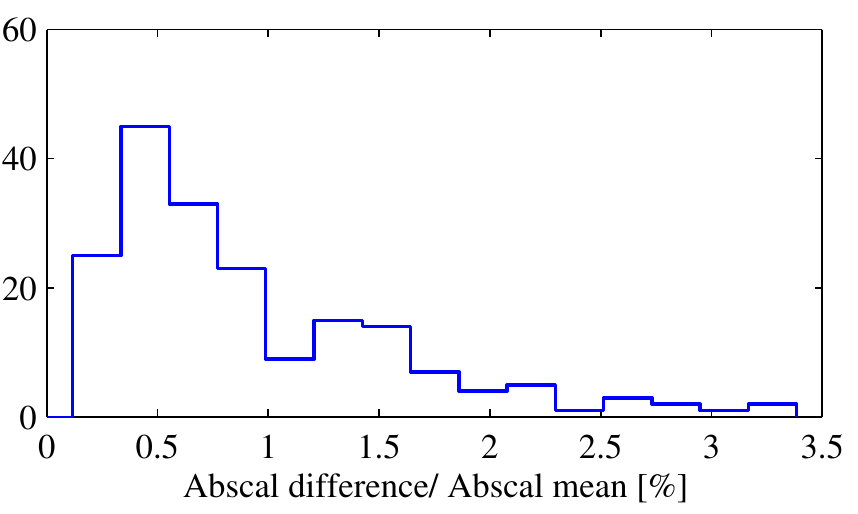}
}
\caption{Spectral response of detectors in the \biceptwo\ camera.  {\it Upper}: histogram of the band centers.  {\it Middle}: histogram of band widths.  {\it Lower}: histogram of absolute calibration (abscal) mismatch, where difference and mean are computed between polarization pairs.  Calibration is through cross-correlation against Plank 143~GHz.  \label{fig:filter_features}}
\end{figure}

We can summarize the spectral response $S(\nu)$ by comparing it to a top-hat response with the same area and then defining the band-center as:

\begin{equation}
\langle \nu \rangle=\int \nu S(\nu) d\nu
\label{eqn:band_center}
\end{equation}
and the bandwidth as
\begin{equation}
\Delta \nu=\frac{(\int S(\nu) d\nu)^2}{\int S^2(\nu) d\nu}.
\label{eqn:band_width}
\end{equation}

The spectral responses of the detectors within a polarization pair must be closely matched to avoid 
systematic contamination through spectral gain mismatch.  In principle, using separate filters for each 
polarization might induce a mismatch through gradients in material properties.  Our design resists this 
by placing the filter pairs in close physical proximity; in the 150~GHz detectors, the filters are $\sim$5~mm apart.  
Furthermore, the filters' magnetic inductance dominates the kinetic inductance; this renders our filters' features 
particularly robust against variation in Nb contamination that might adversely impact electrically long 
resonators~\citep{2005ApPhL..86k4103M}.  As shown in Figure~\ref{fig:filter_features}, our band centers 
and widths are found to be highly repeatable.  
We calibrate our responsivity through correlation against the Planck 143~GHz maps.  Spectral mismatches between filter pairs would manifest themselves as differences in each detector's absolute calibration (abscal), but the bottom panel of Figure~\ref{fig:filter_features} shows that these are less than 3\% of the average responsivity, with a median of 0.5\%.

\subsection{Optical Efficiency}
\begin{figure}[htbp]
\centerline{\includegraphics[width=1\columnwidth]{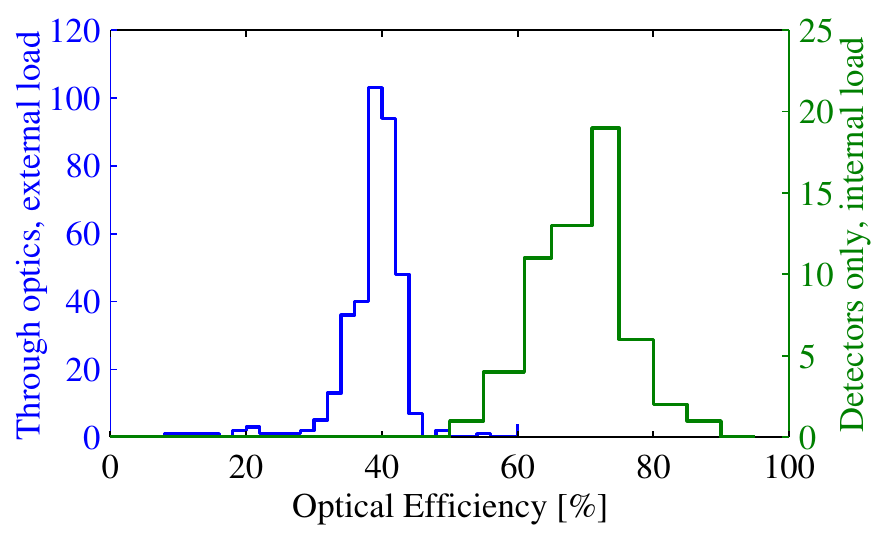}}
\caption{Optical Efficiencies of \biceptwo\ detectors.  Blue curves (left-axis) are end-to-end receiver efficiency through all optics; green curves (right-axis) are raw detector efficiencies for a single test-tile from an engineering-grad test focal plane, in response to an internal cold-load.\label{OE_Keck}}
\end{figure}

We characterize our end-to-end optical efficiency by biasing onto the aluminum TES, which has a higher $T_c$ than Ti, and thus higher saturation power.  By comparing the measured response to aperture-filling 77~K and 300~K sources to the expected incident power $P_{\mathrm{inc}}=k\Delta T \int f(\nu) d \nu$, we can use the  measured spectra to infer efficiency.  We routinely obtain total camera optical efficiencies in excess of 30\%.  For example, the \biceptwo\ efficiencies are shown in Figure~\ref{OE_Keck}.  These figures are for the total camera efficiency, which includes losses other than the detectors.  Measurements of response of early engineering-grade detectors to an internal cold load, which exclude spillover loss and loss in our filter stack, suggest that the raw detector efficiency should be nearly 70\% \citep{oralndo09}.

\section{Sensitivity}
\label{sec:sensitivity}

The array design described above was first deployed for astronomical measurements in 2009 as part of \biceptwo, a ground-based CMB polarimeter.  We have since fabricated dozens of deployment-grade arrays at 95 and 150~GHz for use in the terrestrial \keck\ and balloon-borne \spider\ instruments.  These programs have yielded extensive data on the real-world performance of this technology, as well as demonstrating its adaptability to different optical loads.

\label{sec:noise_model}
\begin{figure}[htbp]
\centerline{\includegraphics[width=1.05\columnwidth]{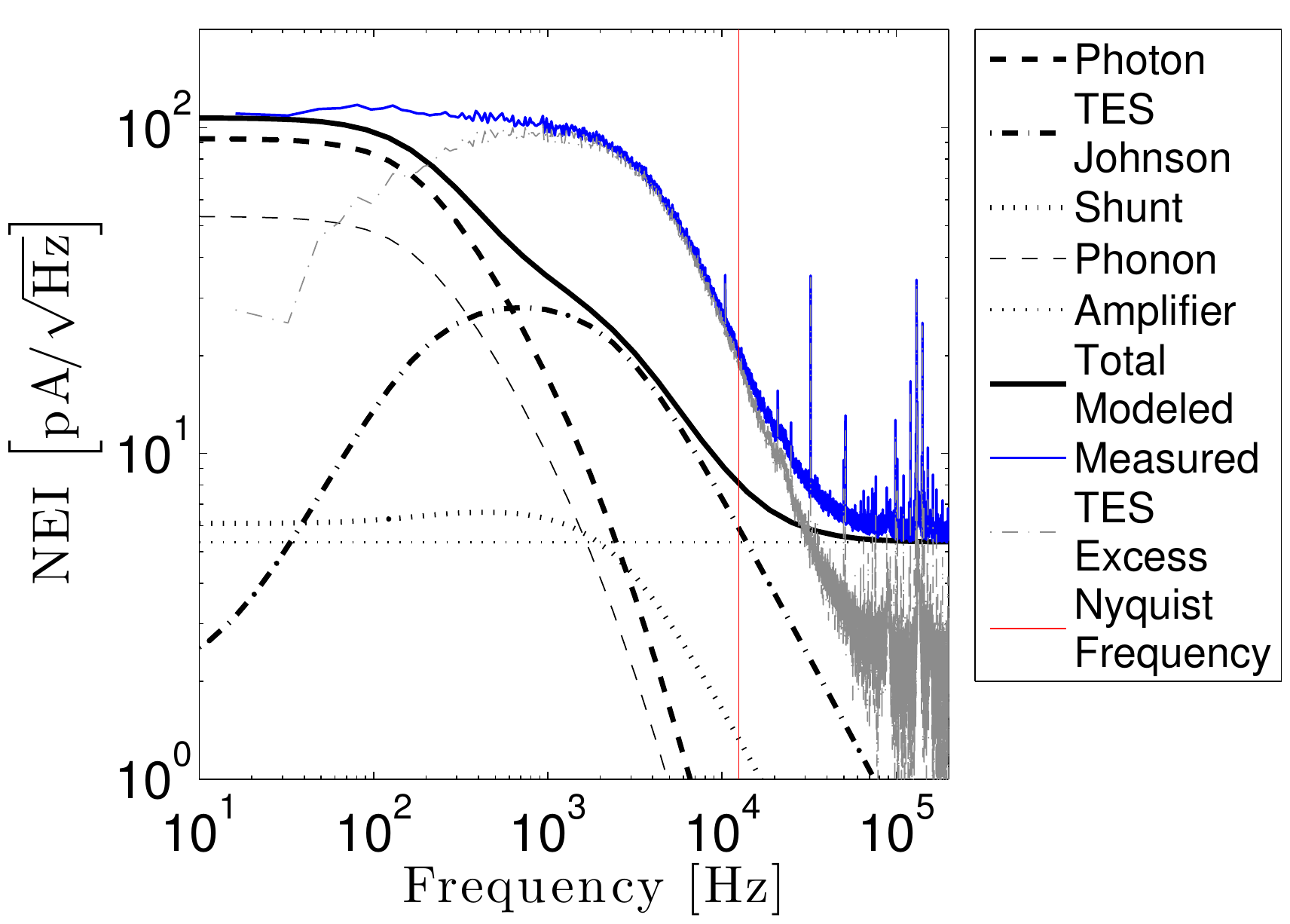}}
\caption{\label{fig:noisemodel}Measured and modeled noise for a single detector in the \keck.  The red line indicates the Nyquist frequency for the multiplexing rate.}
\end{figure}

\subsection{Noise}

The theory of noise in TES bolometers is by now well developed \citep{Irwin_hilton_thy}.  Major contributions typically include photon noise, phonon noise (thermal fluctuation noise), Johnson noise from the TES and shunt resistances, and amplifier noise.  In many cases TESs also exhibit varying degrees of ``excess'' noise beyond that predicted by simplified models.  In multiplexed systems, noise performance also depends upon the relationship between the noise levels and detector and readout bandwidths, since poor choices can lead to substantial noise aliasing (\cite{TDM_MUX}, and \cite{MUX_chips}).  We find that the detectors' measured noise is well reproduced by a simple model incorporating modest excess TES noise, and we have successfully operated the bolometers in configurations with little aliasing penalty.

In order to characterize the noise performance, we have measured noise spectra for the \biceptwo\ and \keck\ detectors using un-multiplexed data digitized at 400~kHz.  Although the frequency range corresponding to degree-scale anisotropies is 0.05--1~Hz for the \biceptwo\ and \keck\ scan strategy, these measurements allow us to observe device performance near the typical multiplexing frequency of 25~kHz and thus model noise aliasing due to the readout.  These measurements, adjusted for expected aliasing, are in good agreement with noise measured in the science-mode multiplexer configuration. 

Figure~\ref{fig:noisemodel} compares the measured noise for a representative detector from the \keck\ to its various modeled components.  Photon noise dominates at the low frequencies of interest for sky observations.  The photon noise equivalent power (NEP) can be expressed as a sum of Bose and shot noise contributions:
  \begin{equation}
    \mathrm{NEP}^2_{\mathrm{photon}}=2h\nu Q_{\mathrm{load}}+\frac{2Q_{\mathrm{load}}^2}{\Delta\nu},
  \end{equation}
where $\nu$ is the band center, $\Delta\nu$ is the bandwidth, and $Q_{\mathrm{load}}$ is the optical loading. The next-largest contribution to the noise at low frequencies is the thermal fluctuation (phonon) noise across the SiN isolation legs, given by Equation~\ref{eqn:Bolo_NEP}.  All other modeled contributions, including the TES Johnson and excess noise, the shunt resistor noise, and the cold and warm amplifiers, are negligible at low frequencies.  

\begin{table*}
\begin{center}
\caption{Summary of detector noise are low frequencies for \biceptwo\ and \keck.}
\begin{tabular}{|l||l|l|l|l|l|l|}
\hline
\rule[-1ex]{0pt}{3.5ex}  Receiver & \biceptwo\ & \keck\ Rx0 & Rx1 & Rx2 & Rx3 & Rx4\\
\hline
\rule[-1ex]{0pt}{3.5ex}  Absorbed Photon Noise [$\mathrm{aW}/\sqrt{\mathrm{Hz}}$] & 41 & 33 & 32 & 34 & 27 & 27\\
\hline
\rule[-1ex]{0pt}{3.5ex}  Phonon Noise [$\mathrm{aW}/\sqrt{\mathrm{Hz}}$] & 27 & 24 & 20 & 24 & 14 & 16\\
\hline
\rule[-1ex]{0pt}{3.5ex}  TES Johnson Noise [$\mathrm{aW}/\sqrt{\mathrm{Hz}}$] & 0.5 & 0.8 & 0.5 & 0.8 & 0.4 & 0.6\\
\hline
\rule[-1ex]{0pt}{3.5ex}  Amplifier Noise [$\mathrm{aW}/\sqrt{\mathrm{Hz}}$] & 2.0 & 2.0 & 2.0 & 2.4 & 2.0 & 2.5\\
\hline
\rule[-1ex]{0pt}{3.5ex}  Total Noise + Aliased at 25~kHz [$\mathrm{aW}/\sqrt{\mathrm{Hz}}$] & 56 & 46 & 39 & 48 & 36 & 40\\
\hline
\rule[-1ex]{0pt}{3.5ex}  per-detector NET from maps  [$\mu\mathrm{K}_{\mathrm{CMB}}\sqrt{\mathrm{s}}$] & 305 & 316 & 351 & 317 & 370 & 429\\
\hline
\end{tabular}
\label{tab:noisemodel}
\end{center}
\end{table*}

At frequencies of $\approx 1~\mathrm{kHz}$ the TES excess noise starts to contribute
significantly.  The TES excess noise, as described in \citep{Gildemeister_excess_noise}, tends to be
proportional to the TES transition steepness $\beta=\left(R/I\right)\left(\partial I/\partial R\right)|_{T}$,
which for these detectors is higher at lower resistances.  The excess noise varies between fabrication batches, although the detector in Figure~\ref{fig:noisemodel} has a relatively low amount. The excess noise should have little contribution to an experiment's overall noise level as long as it is not
aliased.  \biceptwo\ and the \keck\ avoid such aliasing by multiplexing at 25~kHz
and biasing at relatively high resistances on detectors that have large excess noise components.  \spider\ cannot 
multiplex as quickly due to long cable lengths, so we use a lower TES resistance  ($R_n\sim30$~m$\Omega$) and higher Nyquist inductance (2.0~$\mu$H) to limit aliasing of excess noise.

The noise contributions for \biceptwo\ and the \keck\ at low frequencies and under
winter atmospheric loading conditions are broken down in Table~\ref{tab:noisemodel} for the observing year 2012.  The photon noise is of absorbed power only to aid the comparison between models and measurements without adding an extra factor of optical efficiency.  The last line of the table shows total NET computed from jackknife maps, which necessarily references noise of the incident photons and accounts for the optical efficiency of the cameras.

\subsection{Performance of ground-based designs}
\label{sec:B2_Keck}
\begin{figure}[htbp]
\centerline{\includegraphics[width=1.05\columnwidth]{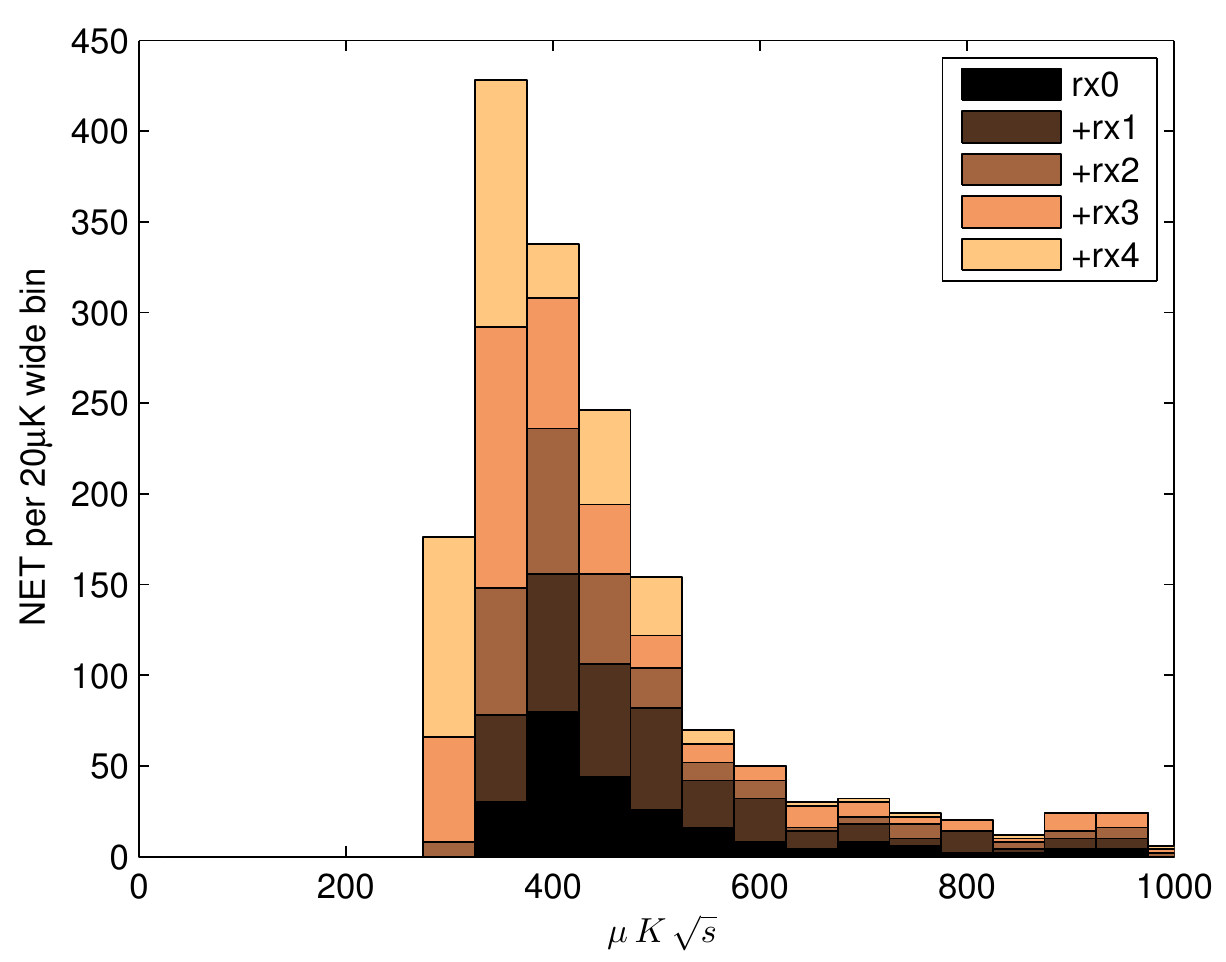}}
\caption{\label{fig:netperdet}NET per detector histogram for the \keck\ in 2013}
\end{figure}

We have measured the devices' performance on the sky during lengthy
\biceptwo\ and \keck\ observation campaigns.  The sensitivity of
the experiment can be measured using the timestream noise between
0.1--1~Hz, after calibrating to the CMB.  \biceptwo\ had a noise equivalent
temperature (NET) of 16~$\mu \mathrm{K}_\mathrm{CMB} \sqrt{\mathrm{s}}$.  The larger detector
complement of the \keck\ achieved 11.5 and 9~$\mu \mathrm{K}_\mathrm{CMB} \sqrt{\mathrm{s}}$
in 2012 and 2013, respectively, also at 150~GHz.  Figure~\ref{fig:netperdet} shows the
distribution of per-detector sensitivity for the \keck\ in 2013.

Another measure of instrument performance is the total map depth achieved,
defined as the noise within a map pixel of a specific size.  This is a function
of instrumental sensitivity, scan strategy, and observing time.  Over three seasons of observing, \biceptwo\ achieved an rms noise of 
$\mathrm{NEQ}=87~\mathrm{nK}\cdot \mathrm{deg}$  ($5.2~\mu\mathrm{K}\cdot\mathrm{arcmin}$).   (\cite{b2instrument}, \cite{biceptwoI}).  Over two years, the five \keck\ 150~GHz cameras achieved $\mathrm{NEQ}=74~\mathrm{nK}\cdot \mathrm{deg}$  ($4.4~\mu\mathrm{K}\cdot\mathrm{arcmin}$).  Combining \biceptwo\ and \keck\ and averaging across the entire effective area 400 square degree field results in rms noise of $2~\mathrm{nK}$.

\section{Conclusions and future work}
\label{sec:conclusions}

We have demonstrated our novel detectors' performance through 18 camera-years of observations in the \biceptwo/\keck\ program, including a detection reported in 2014 of degree-scale B-mode anisotropy \citep{biceptwoI}.
This paper has described the design principles, challenges, fabrication techniques and our characterization/screening program that made these deployments possible.  This program has continued through the recent, successful 2015 \spider\ flight and higher frequency upgrades to \keck.  

We have also deployed \bicepthree\ in the current (2014-2015) Antarctic summer season, which has 1152 95~GHz detectors in a single camera.  Once all the detectors are installed in the 2015-2016 season, this instrument will support 2560 95~GHz detectors.  We package these detectors in individual modules that efficiently fill the focal plane and we illuminate the detectors' antennas with non-uniform Gaussian tapered slot illuminations to reduce spillover onto the camera's stop.  These recent modifications will be the subject of a future paper.

Lastly, our detector technology allows for multi-color focal planes where different color channels are co-located on the focal plane.  Most competing dual-band detector technologies use a single common aperture for all color channels and thus there is a reduction in per-detector efficiency due to aperture spillover \citep{2013ApPhL.102f3506O}.  Planar antenna arrays allow each color to have a custom aperture, thus more efficiently using of both focal plane real estate and limited readout capacity.  
These are under active development for \bicepthree\, further upgrades to the \bicep/\keck~program, and for future \spider\ flights.

\acknowledgements
The development of antenna-coupled detector technology was supported by the JPL Research and Technology
Development Fund and grants 06-ARPA206-0040 and 10-SAT10-0017
from the NASA APRA and SAT programs.  The development and testing of
focal planes were supported by the Gordon and Betty Moore Foundation
at Caltech.  Readout electronics were supported by a Canada Foundation
for Innovation grant to UBC.  The receiver development was supported
in part by a grant from the W. M. Keck Foundation.
\bicep2 was supported by the US National Science Foundation under
grants ANT-0742818 and ANT-1044978 (Caltech/Harvard) and ANT-0742592
and ANT-1110087 (Chicago/Minnesota).  
The computations in this paper were run on the Odyssey cluster
supported by the FAS Science Division Research Computing Group at
Harvard University.  Tireless administrative support was provided by
Irene Coyle and Kathy Deniston.

\pagebreak

\bibliographystyle{apj}
\bibliography{det_bib}

\end{document}